\documentclass{mn2e}
\usepackage{times}
\usepackage{amsmath}
\usepackage{amssymb}

\usepackage{natbib}
\usepackage{epsf}
\usepackage{graphicx}

\newcommand{\etal}{{\it et al.}}            

\newcommand{\nh}{$N_{\rm H}$}

\newcommand{\XMM}{{\it XMM-Newton }}

\newcommand{\Ch}{{\it Chandra }}

\title{{\it{Chandra}} and {\it{XMM-Newton}} observations of NGC~6251}
\author[D. A. Evans et al.]
       {D. A. Evans$^1$\thanks{Email: D.A.Evans@bristol.ac.uk},
        M. J. Hardcastle$^{1,2}$, 
        J. H. Croston$^{1,3}$,
        D. M. Worrall$^1$,
        and M. Birkinshaw$^1$\\
        $^1$ Department of Physics, University of Bristol, Royal Fort,
	Tyndall Avenue, Bristol BS8 1TL, UK\\
$^2$ School of Physics, Astronomy \& Mathematics, University of
	Hertfordshire, College Lane, Hatfield AL10 9AB, UK\\
$^3$ Service d'Astrophysique, CEA Saclay, L'Orme des Merisiers, 91191 Gif-sur-Yvette, France}

\begin{document}

\date{Accepted 2005 February 3. Received 2005 January 28; in original form 2004 October 15}

\maketitle

\label{firstpage}

\begin{abstract}

We present new X-ray observations of the nucleus, jet and extended
emission of the nearby radio galaxy NGC~6251 using the {\it
Chandra}/ACIS-S camera, together with a reanalysis of archival {\it
Chandra}/ACIS-I and {\it XMM-Newton}/EPIC data. We find that the
nuclear X-ray spectrum is well-fitted with an absorbed power-law, and
that there is tentative, but not highly significant, evidence for Fe
K$\alpha$ emission. We argue that the observed nuclear X-ray emission
is likely to originate in a relativistic jet, based on the
double-peaked nature, and our synchrotron self-Compton modelling, of
the radio-to-X-ray spectral energy distribution. However, we cannot
rule out a contribution from an accretion flow. We resolve X-ray jet
emission in three distinct regions, and argue in favour of a
synchrotron origin for all three; inverse-Compton emission models are
possible but require extreme parameters. We detect thermal emission on
both galaxy and group scales, and demonstrate that hot gas can confine
the jet, particularly if relativistic beaming is important. We show
evidence that the radio lobe has evacuated a cavity in the
X-ray-emitting gas, and suggest that the lobe is close to the plane of
the sky, with the jet entering the lobe close to the surface nearest
to the observer.

\end{abstract}


\section{Introduction}

NGC~6251 is a nearby ($z$ = 0.0244) Fanaroff Riley type I (FRI) radio
galaxy with a supermassive black hole of mass (4--8)$\times$10\(^{8}\)
M$_\odot$ (\citealt{fer99}). It exhibits complex morphology on a range
of angular scales, and shows a rich variety of structure across the
electromagnetic spectrum. Optically, NGC~6251 hosts a well-defined
dust disk of diameter 730 pc for (H$_0$ = 70 km s$^{-1}$ Mpc$^{-1}$)
(\citealt{fer99}) and, based on velocity widths of allowed and
forbidden lines, has been classified as a Seyfert 2 galaxy
(\citealt{shu81}). NGC~6251 is the parent galaxy of a spectacular
radio jet, which extends 4.5 arcmin (130 kpc) north-west (NW) of the
nucleus and has an opening angle of 7.4$^\circ$
(\citealt{wag77,per84}). On smaller (milliarcsecond) scales, VLBI
observations (e.g.,
\citealt{jonetal86}) have detected radio emission from an unresolved
core, and a parsec-scale jet extending to the NW of the nucleus. A
detection of a parsec-scale counterjet was claimed by
\cite{sud00}, but disputed by \cite{jon02}, who placed a lower limit
of 128 on the jet/counterjet brightness ratio, implying an angle to
the line of sight $\theta < 40^\circ$. VLA spectral-line radio
observations of the inner regions of NGC~6251 (\citealt{wer02}) show
21-cm neutral hydrogen absorption, corresponding to an intrinsic
column of \nh = (4.5$^{+2.6}_{-2.4}$)$\times$10\(^{20}\) atoms
cm$^{-2}$. At higher energies, it has been claimed that NGC~6251 is
associated with the {\it{CGRO}}/EGRET gamma-ray source 3EG J1621 + 8203
(\citealt{muk02,fos04}).

X-ray emission from NGC~6251 was first detected with {\it ROSAT}
(\citealt{bw93}), and has more recently been imaged in detail with \Ch
and {\it{XMM-Newton}}. \cite{bw93} presented {\it ROSAT} PSPC
observations of NGC~6251, detecting nuclear emission and a
contribution from a hot, X-ray emitting halo. They argued that the
kiloparsec-scale radio jet cannot be confined by gas pressure
alone. \cite{mac97a} reanalysed the {\it ROSAT} data, and confirmed
the presence of an elliptically-shaped thermal X-ray halo extending
out to $\sim$ 100 kpc, reporting enhanced X-ray emission coincident
with components of the radio jet on 100-kpc scales. X-ray emission
from the inner jet of NGC~6251 was imaged by \cite{har99} using the
{\it ROSAT} HRI. \cite{ker03} used a {\it{Chandra}}/ACIS-I observation
of NGC~6251 to argue that the pressure of the extended halo is
sufficient to confine the radio jet. Their observation also confirmed
the PSPC detection of the outer X-ray components of the jet. A
subsequent {\it XMM-Newton} EPIC observation of the halo
(\citealt{sam04}) measured a temperature for the gas of $kT$ = 1.6
keV. The \XMM spectrum of the outer regions of the X-ray jet ruled out
a thermal model for the X-radiation, and \cite{sam04} argued that the
emission was due to beamed inverse-Compton scattering of cosmic
microwave background (CMB) photons to X-ray energies.

The physical origin of the nuclear X-ray continuum of NGC~6251, as
with other radio galaxies, is uncertain. Two popular interpretations
for the origin of the emission are: (1) accretion phenomena in or near
a disk; and (2) nonthermal processes (inverse-Compton or synchrotron
emission) in a subparsec-scale jet. The first model is often supported
by the detection of high equivalent width Fe K$\alpha$ lines. The
second interpretation is supported for radio galaxies in general by
the observed correlation between the luminosities of soft, unresolved
X-ray emission and 5-GHz core radio emission (e.g.,
\citealt{wor94,can99,har99}). The nuclear spectrum of NGC~6251, as
measured with {\it ASCA} (\citealt{tur97}) and \XMM (\citealt{gli04}),
was modelled with a power law and a high equivalent width fluorescent
Fe K$\alpha$ emission line, suggesting that the X-ray emission
originates from an accretion flow. However, the power-law continuum
measured with BeppoSAX (\citealt{gua03}) had a somewhat flatter photon
index and higher luminosity than the earlier {\it ASCA} observation in
particular, with no Fe K$\alpha$ line detected. Moreover, the
double-peaked SED measured by, for example,
\cite{gua03}, \cite{ho99}, and \cite{chi03}, led those authors to
suggest that the dominant X-ray-production mechanism was synchrotron
self-Compton (SSC) emission from a jet.

This paper is organised as follows. Section 2 contains a description
of the data and a summary of our analysis. The results from our
analysis of the nucleus, jet, and extended emission are presented in
Section 3, and are discussed in Section 4. We end with our conclusions
in Section 5. All results presented in this paper use a cosmology in
which $\Omega_{\rm m, 0}$ = 0.3 and $\Omega_{\rm \Lambda, 0}$ = 0.7,
and H$_0$ = 70 km s$^{-1}$ Mpc$^{-1}$. At the redshift of NGC~6251
($z$ = 0.0244), 1 arcsec corresponds to 485 pc . All spectral fits
include absorption through our Galaxy using $N_{\rm H, Gal}$ =
5.82$\times$10\(^{20}\) atoms cm$^{-2}$ (\citealt{mur96}). When
distinguishing between different model fits to the data, we present
$F$-statistic results for ease of comparison with other papers,
although we note that this method may be unreliable in such
circumstances (\citealt{pro02}). We adopt thresholds of 95 and 99.9
per cent for marginally and highly significant improvements in the
fit, respectively. Errors quoted in this paper are 90 per cent
confidence for one parameter of interest (i.e., $\chi^2_{\rm min}$ +
2.7), unless otherwise stated.

\section{Observations and Analysis Methods}

\subsection{Chandra}
\label{chandra-obs}

NGC~6251 was observed on 2003 November 11 (ObsID 4130) with the S3 and
S4 chips of the {\it{Chandra}}/ACIS charge-coupled devide (CCD)
camera. The observation was made in the VFAINT mode, using a 128-row
subarray (giving a $1 \times 16$ arcmin field of view), in order to
reduce the CCD frame time to 0.54 s and lessen the effect of
pileup. The data were reprocessed using {\it{CIAO v3.0.2}} with the
{\it{CALDB}} v2.26 calibration database to create a new level-2 events
file with grades 0, 2, 3, 4, 6, afterglow events preserved, and the
0.5-arcsec pixel randomization removed. To check for intervals of high
particle background, light curves were extracted for the ACIS-S3 chip,
excluding the core. Inspection of the light curves revealed the
presence of several flares which were removed from our analysis,
reducing the total exposure time from 46.5 ks to 43.0 ks.

The \Ch observation of the nucleus of NGC~6251 is piled up. We
demonstrate this using two independent methods. First, the counts
extracted from a source-centred circle of radius 1.23 arcsec (2.5
pixels) in the 0.5--7 keV energy band gave 0.35 counts per frame
exposure time, for which a pileup fraction of $\sim 10$--15 per cent
is predicted using the {\sc{PIMMS}} software. As a second independent
test, we estimated the spatial extent of the pileup by producing an
image consisting of `afterglow' events, which are produced from
residual charges from either cosmic-ray events or from piled-up
events. We computed the numbers of afterglow events per unit area
within two extraction regions: (1) a source-centred circle of radius
0.492 arcsec (1 pixel); and (2) an annulus of inner radius 0.492
arcsec (1 pixel) and outer radius 1.23 arcsec (2.5 pixels). The ratio
of these numbers was found to be 55 $\pm$ 6. We performed a similar
test with 2--5 keV events in the same regions, and found the ratio to
be 19 $\pm$ 1, showing that the effect of pileup is reduced in our
annular extraction region. The techniques we used to measure the core
spectrum in the presence of this pileup are discussed in Section
\ref{chcore}.

In addition to the new ACIS-S data, we made use of an earlier 40-ks
ACIS-I observation taken on 2000 September 11 (ObsID 847). These data
were discussed by \cite{ker03} and \cite{sam04}.  Because the X-ray
core and inner jet were placed on the chip gaps in this observation,
the data are not useful in constraining the properties of the inner
regions of the source, but they provide additional information on the
large-scale jet and extended emission.

\Ch images presented in this paper were made using the 0.5--5.0 keV 
counts in the filtered ACIS-S dataset. For the analysis of extended
emission, we fitted radial surface-brightness profiles with models
convolved with the \Ch point spread function (PSF), and extracted physical parameters
(\citealt{bw93}). In the \Ch analysis of the core, the centroids of
all circular and annular extraction regions were determined using the
{\sc zhtools} software (A.  Vikhlinin, private communication).

\subsection{XMM-Newton}
\label{xmm}

The {\it XMM-Newton} observation of NGC~6251 (\citealt{gli04,sam04})
was extracted from the archive for comparison with the \Ch datasets,
and in particular to study the emission from the group environment and
the large-scale jet. The data were reprocessed using the latest
software and calibration files available from the \XMM project, {\sc{SAS}}
version 6.0.0. The total exposure time was 49.3 ks (MOS1), 49.4 ks
(MOS2), and 41.1 ks (pn).

\begin{figure*}
\begin{center}
\includegraphics[width=16cm]{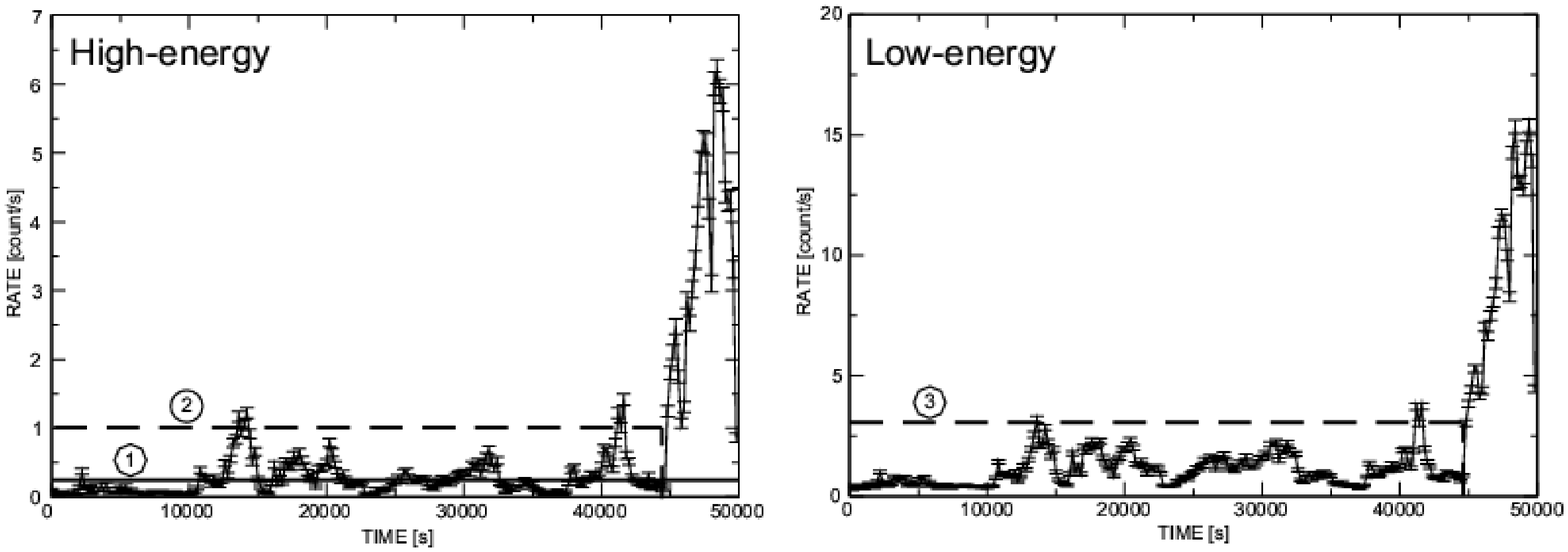}
\caption{\XMM MOS1 light curves used to evaluate good-time intervals. Left: High-energy (10-12 keV) light curve for the whole field of view. Right: Low-energy (0.4--10 keV) light curve for CCD 1, excluding the source. Also shown are the time/count rate filtering criteria in the conservative case (1) and the $3\sigma$ filtering case (2 and 3), as defined in Section 3.2.}\label{mos1_ref_ref2_histo_sup}
\end{center}
\end{figure*}

To examine any periods of high background, we extracted high-energy
light curves for the MOS1, MOS2, and pn cameras, in the energy range
10--12 keV (MOS) and 12--14 keV (pn), using the entire field of view
in each case. The data were filtered to include events with the
PATTERN=0 and FLAG=\#XMMEA\_EM (MOS) and \#XMMEA\_EP (pn) attributes
only. For a more direct comparison with the method of \cite{gli04}, we
extracted low-energy light curves for each camera in the energy range
0.4--10 keV, using only the CCD upon which the source was located
[CCD\_ID=1 (MOS); CCD\_ID=4 (pn)], but excluding a circle of radius 35
arcsec centred on the source. PATTERN $\leq 12$ (MOS) and $\leq 4$
(pn), together with FLAG=0 attributes were selected.  Inspection of
both the high-energy and low-energy light curves (see
Fig.~\ref{mos1_ref_ref2_histo_sup}) revealed several periods of high
background, which we shall discuss further in Section
\ref{backgnd}. We also note from Figure~\ref{mos1_ref_ref2_histo_sup}
that there is little difference between the location and amplitudes of
the flares extracted using the high- and low-energy light curve
methods.

The \XMM observation of the nucleus does not suffer from significant
pileup. The count rates for a source-centred circular extraction
region of radius 35 arcsec were found for each of the MOS1, MOS2, and
pn cameras, and are given in Table~\ref{xmm_pileup_tab}, together with
the nominal count rates at which pileup is expected to become
important. In addition, we followed the prescription of \cite{mol03}
to produce an image of diagonal bipixels in the MOS cameras (assigned
PATTERN numbers 26--29), which are produced almost solely by the
pile-up of two single-pixel events. No such events were
found. 

Spectral analysis on the large-scale jet and extended emission regions
was performed using scripts based on the {\sc sas} {\sc evselect} tool
to extract spectra from all three cameras. We included the vignetting
correction by using the {\sc sas} task {\sc evigweight}, and therefore
used on-axis response files created using the {\sc sas} tasks {\sc
rmfgen} and {\sc arfgen} (for a detailed description of this method
see \citealt{arn02}). For the jet regions, we used local background
subtraction in order to remove thermal emission from the surrounding
hot gas. For the extended emission, however, it was necessary to use a
double subtraction technique (e.g. \citealt{arn02}) both for spectral
analysis and for the radial surface brightness profiling used to study
the spatial distribution of the emission. The double subtraction
technique first takes account of the high level of particle background
(which the vignetting corrections incorrectly weights up at large
off-axis angles) by subtracting the counts from identical regions of a
background template file, scaled to account for any differences in the
particle background level of the source and background datasets. For
this step, we used the background template files of \cite{rea03}
appropriate for each instrument and filter, with the vignetting
correction applied in the same way as for the source events lists. The
second step of the process accounts for the residual background by
subtracting a local background taken from an outer region free of
source emission. This step is necessary first because the level of
X-ray background events may be different in the source and background
datasets, as a result of different levels of Galactic absorption and cosmic
variance, and also because the scaling required to correctly remove
the particle background introduces an arbitrary scaling factor into
the other components of the background that must be taken into
account. For further details of the technique, see \cite{arn02}, and
for more discussion of our background analysis methods for \XMM data,
see \cite{cro03}.

The radial surface brightness profiles (obtained separately for each
\XMM camera) were analysed by fitting point-source-only models and
convolved point-source plus $\beta$-models to determine the
significance of the contribution from extended emission. We modelled
the \XMM PSF using the most recent CCF (Current Calibration File)
components: files XRT1\_PSF\_0007.CCF, XRT2\_PSF\_0007.CCF and
XRT3\_PSF\_0006.CCF, described in the document CAL-SRN-0167-1-1,
available from the \XMM website\footnote{http://xmm.vilspa.esa.es}.

Images were produced from the \XMM data using the techniques described
in \cite{cro03} to interpolate over the chip gaps and remove
contaminating background point sources (identified using the {\it
{Chandra}}/ACIS-I data and an adaptively smoothed image of the \XMM data
for larger radii). The image was not vignetting corrected, as the
effect of incorrectly weighting up particle events at large radii
would dominate over any real emission. This means that structure at
larger radii is revealed in our image, but that its surface brightness
is underestimated. The resulting image was smoothed using Gaussian
kernels to show the distribution of extended emission, and also
adaptively smoothed to show compact structure in the regions of
extended emission.
\label{xmmimaging}

\subsection{Radio}

The X-ray and radio structures were compared using a number of radio
datasets. The best low-resolution image available was the map provided
by the 3CRR Atlas\footnote{http://www.jb.man.ac.uk/atlas/}, a 327-MHz
Westerbork Synthesis Radio Telescope (WSRT) image first published by
\cite{mac97b}. This image has a resolution (FWHM of restoring
circular Gaussian) of 55 arcsec. On smaller scales, we used a variety
of NRAO Very Large Array (VLA) datasets, including the 1.4-GHz map of
\cite{wer02}, 330-MHz maps taken for other purposes by one of us
(Birkinshaw et al., in preparation) and datasets at 1.4, 1.6 and 8.4
GHz taken from the VLA archive, including the data used to make the
images presented by \cite{jonetal86} and \cite{jon94}. A full list of
the VLA datasets used and their properties is given in
Table~\ref{vla}. Where we obtained VLA data from the archive we
reduced them in the standard manner within {\sc{AIPS}}, calibrating the flux
with respect to a standard source, usually 3C\,286. Maps made by
others were reduced in a similar way. We combined the 1.6-GHz data of
\cite{jonetal86} and \cite{jon94}, which are the most sensitive
datasets available to us, subtracting from the A-configuration dataset
to account for the effects of core variability, to obtain a combined
dataset with sensitivity and resolution somewhat higher than that of
the early maps of \cite{per84}. This dataset is used in several of the
images shown in this paper. In addition, high-resolution VLBI maps were made
available to us (D. Jones, private communication). We discuss these in
Section \ref{vlbi}.
\label{radiomaps}

\section{Results}

\subsection{Overview of X-ray properties of NGC~6251}

\begin{figure*}
\begin{center}
\includegraphics[width=16cm]{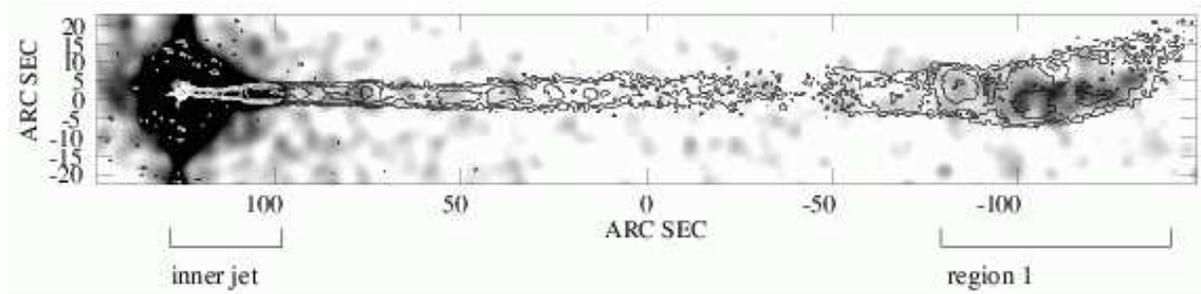}
\caption{The inner jet and region 1 of the outer jet (as defined in
    the text) of NGC 6251 as seen in the new {\it Chandra}/ACIS-S data.
    The image is smoothed with a 4-arcsec FWHM circular Gaussian;
    superposed are contours of the 1.6-GHz radio map discussed in
    Section \ref{radiomaps} at $0.2 \times (1,2,4\dots)$ mJy beam$^{-1}$. The
    image shown here has been rotated through $-24.4$ degrees.}
\label{wholej}
\end{center}
\end{figure*}

\begin{figure*}
\begin{center}
\includegraphics[width=10cm]{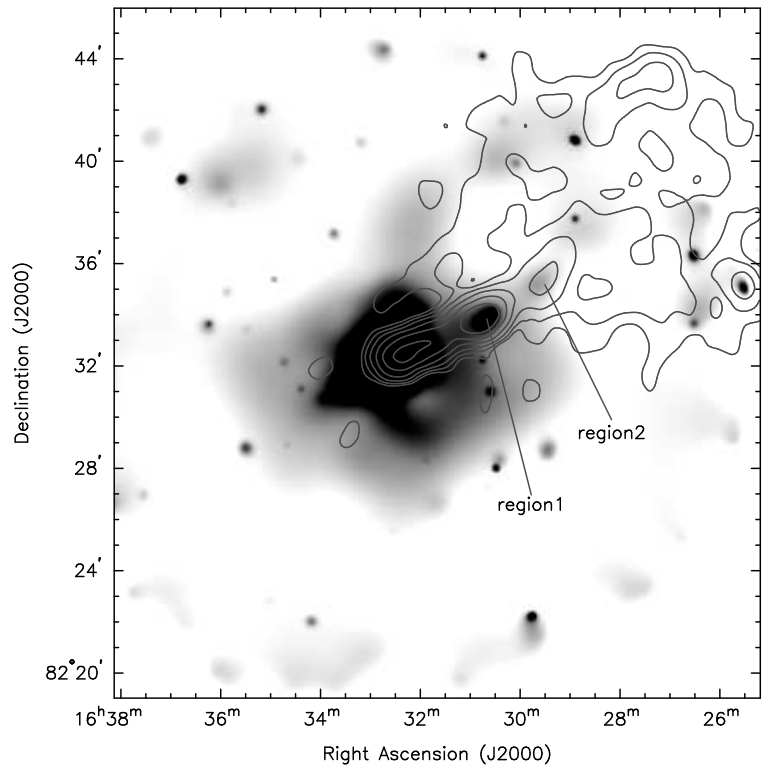}
\caption{Adaptively smoothed, combined MOS1, MOS2 and pn image in the
  0.5--5 keV energy band,
  with radio contours overlaid. The image shows the relation
  between the X-ray and radio jet and large-scale extended emission.
  Radio contours are from the Westerbork 327-MHz map discussed in
  Section \ref{radiomaps}. Contour levels
  are 1,2,4...16 $\times 2 \times 10^{-2}$ Jy beam$^{-1}$. The
  radio/X-ray source on the W edge of the image is presumably a
  background AGN, though no optical counterpart is detected in DSS images.}
\label{ext_overlays}
\end{center}
\end{figure*}

Figures~\ref{wholej} and~\ref{ext_overlays} show the X-ray emission
from NGC~6251 detected with {\it Chandra}/ACIS-S and
{\it{XMM-Newton}}, respectively, with radio contours overlaid. We
divide the X-rays from NGC~6251 into emission from the nucleus
(Section 3.2), emission from the jet (Section 3.3), and extended
thermal emission (Section 3.4). The nucleus, the brightest component,
is detected with both \Ch and {\it XMM-Newton}. Components of the jet
are also detected with both telescopes, although our new {\it
Chandra}/ACIS-S observation is required to give a good image of the
inner 15 arcsec of the jet. Extended thermal X-ray emission is
observed on the galaxy scale with {\it Chandra}/ACIS-I and ACIS-S
(Fig.~\ref{wholej}), and on the group scale with {\it Chandra}/ACIS-I
and \XMM (Fig.~\ref{ext_overlays}) (see Section 3.4).

\subsection{Core}

\subsubsection{Spectrum}
\label{chcore}

a) \Ch

The {\it Chandra}/ACIS-S spectral fitting was performed over an energy
range 0.5--7 keV. We chose the lower bound of the energy range to be
0.5 keV, because of the known degradation of the ACIS quantum
efficiency energies below $\sim$ 0.4--0.5 keV (\citealt{mar04}). Owing
to the pileup known to be present, we tried several methods to extract
and model the core spectrum. The first approach was to extract the
spectrum using a source-centred circular extraction region of radius
1.23 arcsec (2.5 pixels), and to model the effect of the strong pileup
known to be present using the method of \cite{dav01} as implemented in
{\sc{Sherpa}}.

The second approach was to follow \cite{gam03} and reduce
complications due to pileup by extracting the spectrum using an
annulus surrounding the source, sampling the wings of the PSF and
excluding the most heavily piled-up region. The inner radius of the
annulus was chosen to be 0.492 arcsec (1 pixel), which encloses
$\sim$60 per cent of the core flux. We chose the outer radius to be
1.23 arcsec (2.5 pixels), which encloses $\sim 90$--95 per cent of the
core flux.  The corresponding point-like ARFs (ancillary response files)
were corrected for the energy-dependent missing flux using
software (M. Tsujimoto, private communication) that calculates the
encircled energy fraction in an annular extraction region of a model
PSF created using ChaRT and MARX. This theoretical PSF was smoothed by
0.35 arcsec to take into account the broadening of the PSF associated
with aspect and pixelization. Our spectral fits using this method were
consistent with those in which the ARF was simply scaled by the
fraction of the PSF missing in the annulus (i.e., when the energy
dependence was not taken into account). Since pileup could still
affect the spectrum of the annulus, we compared the results of
modelling the spectrum both with and without implementing the
pileup method of \cite{dav01}.

We compare both methods here. The spectral parameters obtained from
the 2.5-pixel circle using the pileup-model [$\Gamma = 1.83 \pm 0.06$,
normalization = $(1.49 \pm 0.04) \times 10^{-3}$] method were
inconsistent with those from the annulus method [$\Gamma = 1.61 \pm
0.03$, normalization = $(0.87 \pm 0.02) \times 10^{-3}$]. When the
effect of the moderate pileup present in the annulus was taken into
account with the \cite{dav01} method, the spectral parameters [$\Gamma
= 1.72 \pm 0.04$, normalization = $(0.97 \pm 0.03) \times 10^{-3}$]
were consistent to within $\sim 10\%$ of those obtained when the
effect of the pileup was not considered, but still inconsistent with
the results obtained from the circular extraction region.

We decided to investigate this discrepancy further by extracting and
modelling a spectrum from the frame transfer streak, following the
method of \cite{mar05} to simply correct for the fraction of source
events that occurred in the frame transfer streak. We found the
best-fitting photon index, $\Gamma = 1.63^{+0.31}_{-0.28}$ and the
normalization to be $(9.31 \pm 1.90) \times 10^{-4}$. These spectral
parameters are consistent with those found using the annulus
method, but again are inconsistent with those found using the circle
method. We therefore adopted the annulus method for our spectral
analysis, and attribute the discrepancy between the circle and annulus
methods to difficulties with the pileup model in handling data sets
with large pileup fractions. We note in passing that using the annulus
method has the advantage of being able to model spectrally
galaxy-scale thermal emission.

\begin{figure*}
\begin{center}
\includegraphics[width=10cm]{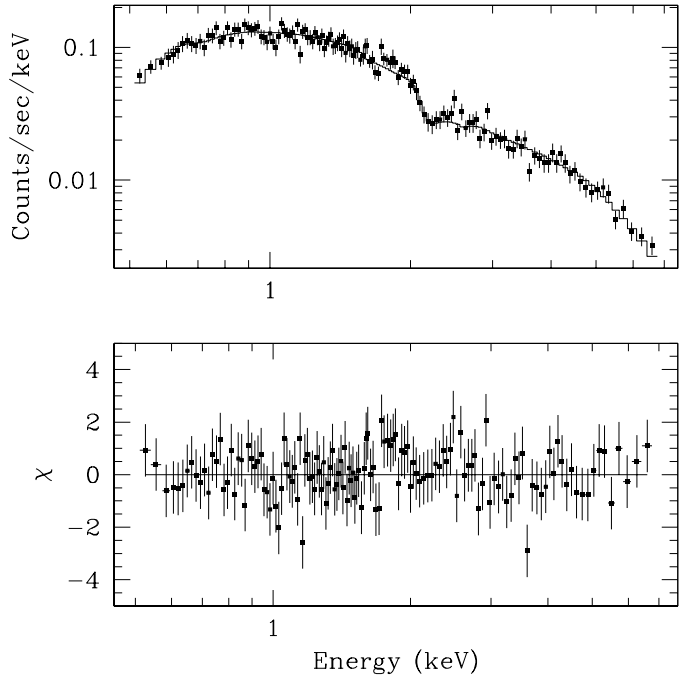}
\caption{Spectral fit to data from the \Ch observation in the energy range 0.5--7 keV with a power-law and {\sc{apec}} model, with the abundance fixed at 0.35 solar. The spectral extraction region was a source-centred annulus of inner radius 0.492 arcsec and outer radius 1.23 arcsec. Contributions to $\chi^2$ are also shown.}
\label{fit6_ref_jdp}
\end{center}
\end{figure*}

With an off-source background extraction region that matched the
region used for radial profiling (see Section 3.4) there were 9,433 net
0.5--7 keV counts in the source-centred annulus. We grouped the data
to 50 counts per PHA bin.  We initially attempted to model the spectrum
with a single, absorbed power-law. The fit was acceptable, but the
intrinsic absorption column was poorly constrained ($N_{\rm H} \sim
8\times 10^{18}$ atoms cm$^{-2}$ with a 90 per cent confidence upper
limit $\sim 3\times 10^{20}$ atoms cm$^{-2}$). This was consistent
with the value measured from 21 cm radio observations [$N_{\rm H} =
(4.5^{+2.6}_{-2.4}) \times 10^{20}$ atoms cm$^{-2}$, \cite{wer02}]. We
therefore decided to fix the absorbing column at its radio-measured
value and performed all subsequent spectral fits using this intrinsic
absorption. An absorbed power law with a photon index $\Gamma = 1.72
\pm 0.04$ gave an acceptable fit ($\chi^2 = 121$ for 136 degrees of
freedom (dof)). We found a significant improvement in the fit with the
addition of a thermal component in the form of a collisionally-ionized
plasma model ({\sc{apec}} in {\sc xspec}), with a temperature $kT =
0.20 \pm 0.08$ keV, and the abundance fixed at 0.35 solar. This model
gave a good fit to the data: $\chi^2 = 107$ for 134 dof, with the
probability of achieving a larger $F$ by chance $\ll$ 0.1 per
cent. The spectrally measured number of 0.5--5 keV counts in the
thermal component between 0.492 and 1.23 arcsec is $99 \pm 28$ ($\pm
46$), in agreement with the number of counts found with the radial
profiling analysis (see Section 3.4), where uncertainties are quoted
as 1$\sigma$ (unbracketed) and 90 per cent (bracketed), for one
interesting parameter. This fit is shown in
Figure~\ref{fit6_ref_jdp}, and the
best-fitting spectral parameters given in
Table~\ref{chandra_xmm_spectral_tab}.

We confirmed the presence of the thermal emission seen by \cite{gli04}
using \XMM data by extracting a spectrum using an annular extraction
region of inner radius 0.492 arcsec and outer radius 35 arcsec (the
same as used by \citealt{gli04}), excluding contaminating emission
from the jet and frame transfer streak. There were 11,319 net counts
in the source extraction region. The best-fitting parameters of the
absorbed power-law model are consistent with those found using the
annulus of inner radius 0.492 arcsec and outer radius 1.23 arcsec, and
the temperature of the thermal component on this scale ($kT = 0.59 \pm
0.05$ keV) is consistent with that quoted by \cite{gli04}. The
measured number of counts in the thermal component between 0.492 and
35 arcsec is $912 \pm 132$ ($\pm 185$), with uncertainties quoted as
above. This number agrees, within the errors, with the number of
thermal counts found in the radial profile analysis.

We searched for the broad 6.4-keV Fe K$\alpha$ emission line reported
by \cite{gli04} in the spectrum extracted from the annulus with inner
radius 0.492 arcsec and outer radius 1.23 arcsec. An absorbed power
law was fitted to the data over an energy range of 2.0--6.0 keV and
6.7--7.0 keV, and the interpolated continuum subtracted from the data
in the energy range 6.0--6.7 keV. The continuum parameters were
consistent with those found previously, and provided a good fit to the
data ($\chi^2 = 8.34$ for 6 dof in the energy range 5.5--7.0 keV). A
Gaussian Fe emission line, with a centroid allowed to vary between 6.2
and 6.7 keV, and free line width was then added to the model. The line
energy was poorly constrained, and the decrease in the fit statistic
($\Delta\chi^2 = 6.18$ for three additional parameters) not
significant on an $F$-test (the probability of a larger $F$ by chance
is 21 per cent). As an additional test, the data were regrouped to 5
counts per PHA bin and the Cash-statistic used to evaluate the
goodness of fit. Again, the line energy was poorly constrained, and
the decrease in fit statistic ($\Delta$C = 9.14) small.

Further, a Gaussian Fe K$\alpha$ emission line, with best-fitting
parameters matching those found with \XMM (\citealt{gli04}), was added
to the \Ch continuum model. The {\it{XMM-Newton}}-measured emission
line gave a decrease in the fit statistic of $\Delta\chi^2 = 3.14$ for
three additional parameters that was not significant (the probability
of a larger $F$ by chance is 65 per cent). Finally, the line
parameters were changed to match those detected with {\it ASCA}
(\cite{tur97}. Again, the decrease in the fit statistic was not
significant ($\Delta\chi^2 = 2.79$ for three additional
parameters). However, we note that the non-detection of Fe K$\alpha$
emission with \Ch is unsurprising, given the relatively low effective
area of the ACIS camera at 6.4 keV.\\

\noindent b) \XMM

A goal of our \XMM analysis was to make an independent evaluation of
the fluorescent Fe K$\alpha$ emission from the nucleus claimed by
\cite{gli04}. As mentioned in Section 2, the \XMM observation of NGC 
6251 is severely contaminated by background flares. Contamination from
background flares is a serious concern in the analysis of the low
surface-brightness jet and extended emission regions, whereas for the
high signal-to-noise core spectra it is less important.  However, it
is still important to assess carefully the effect of the background
flares on the nuclear spectral analysis.

To achieve this, we adopted several good time interval (GTI)-filtering
criteria based upon both the high-energy and low-energy light curves
discussed in Section
\ref{xmm} and presented in Figure~\ref{mos1_ref_ref2_histo_sup}. The criteria we used are described below, and
the effective exposure and number of 4--10 and 5.5--7.5 keV counts
within a 35 arcsec source-centred circle are shown in
Table~\ref{exposures_tab} for each GTI filtering method.
\label{backgnd}
\begin{enumerate}
 
\item Conservative filtering based on 10--12 keV (MOS) and 12--14 keV (pn) light curves. The maximum allowed count rates were 0.25 s$^{-1}$ (MOS) and 1 s$^{-1}$ (pn)
\item $3\sigma$ filtering based on 10--12 keV (MOS) and 12--14 keV (pn) light curves. Good time intervals were created by excluding data after 44.6 ks (MOS) and 42.6 ks (pn) (corresponding to the last flare), and calculating the time periods when the remaining data are within $\pm 3\sigma$ of the mean level.
\item $3\sigma$ filtering based on 0.4--10 keV light curves (MOS and pn). As above.
                                                                                
\end{enumerate}

The \XMM spectral fitting was performed jointly for the MOS1, MOS2 and
pn cameras for an energy range of 0.4--10 keV, using each of the
GTI-filtering methods described above. We extracted source spectra
using a circular extraction region of radius 35 arcsec centred on the
source. Background spectra were extracted using a source-centred
annular extraction region of inner radius 140 arcsec and outer radius
280 arcsec for the MOS cameras, and an off-source rectangle for the pn
camera. Only events with FLAG=0 and PATTERN $\leq$ 12 (MOS) and $\leq$
4 (pn) were included in our analysis. The data were grouped to 40
counts per PHA bin.

\begin{figure*}
\begin{center}
\includegraphics[width=10cm,angle=270]{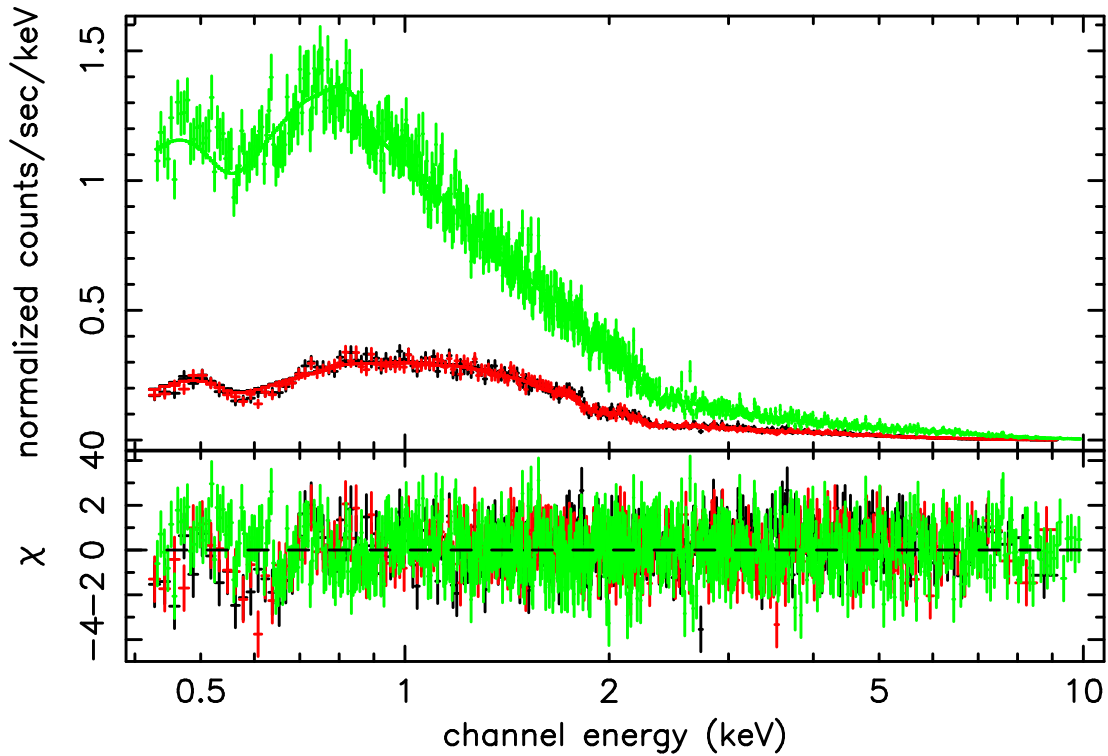}
\caption{Spectral fit to data from the \XMM observation in the energy
  range 0.4--10 keV with a power-law and {\sc{apec}} model, with the
  abundance fixed at 0.35 solar. A Gaussian line was included in the
  fit for completeness. Contributions to $\chi^2$ are shown. The GTI
  filtering was performed using the $3\sigma$ method defined in Section 3.2;
  the continuum parameters found when using the other GTI-filtering
  methods are consistent with this case.}
\label{1pow_apec_finh_z0.35}
\end{center}
\end{figure*}

As a representative example, we describe the fitting to the continuum
performed on the data filtered using the $3\sigma$ method based on the
high-energy light curves data: the spectral parameters derived using
the other GTI-filtering methods are entirely consistent with this
case. We initially attempted to model the spectrum with a single,
absorbed power-law, with the intrinsic absorption was fixed at
$4.5\times 10^{20}$ atoms cm$^{-2}$. The normalizations of all
spectral components were kept free for each camera. For completeness,
and following \cite{gli04}, an Fe K$\alpha$ emission line at $\sim$
6.4 keV was modelled with a Gaussian function and is discussed later.
The fit was poor: $\chi^2 = 1347$ for 1135 dof. However, a substantial
improvement in the fit was obtained by adding a thermal component with
a temperature $kT = 0.60 \pm 0.07$ keV, and abundance fixed at 0.35
solar. The new fit was acceptable ($\chi^2 = 1197$ for 1131 dof), with
the probability of achieving a larger $F$ by chance $\ll 0.1$ per
cent. The counts spectrum and best-fitting model, together with
contributions to $\chi^2$ is shown in
Figure~\ref{1pow_apec_finh_z0.35}. The best-fitting spectral
parameters for the \XMM observation are given in
Table~\ref{chandra_xmm_spectral_tab}.

\begin{figure*}
\begin{center}
\includegraphics[width=10cm,angle=0]{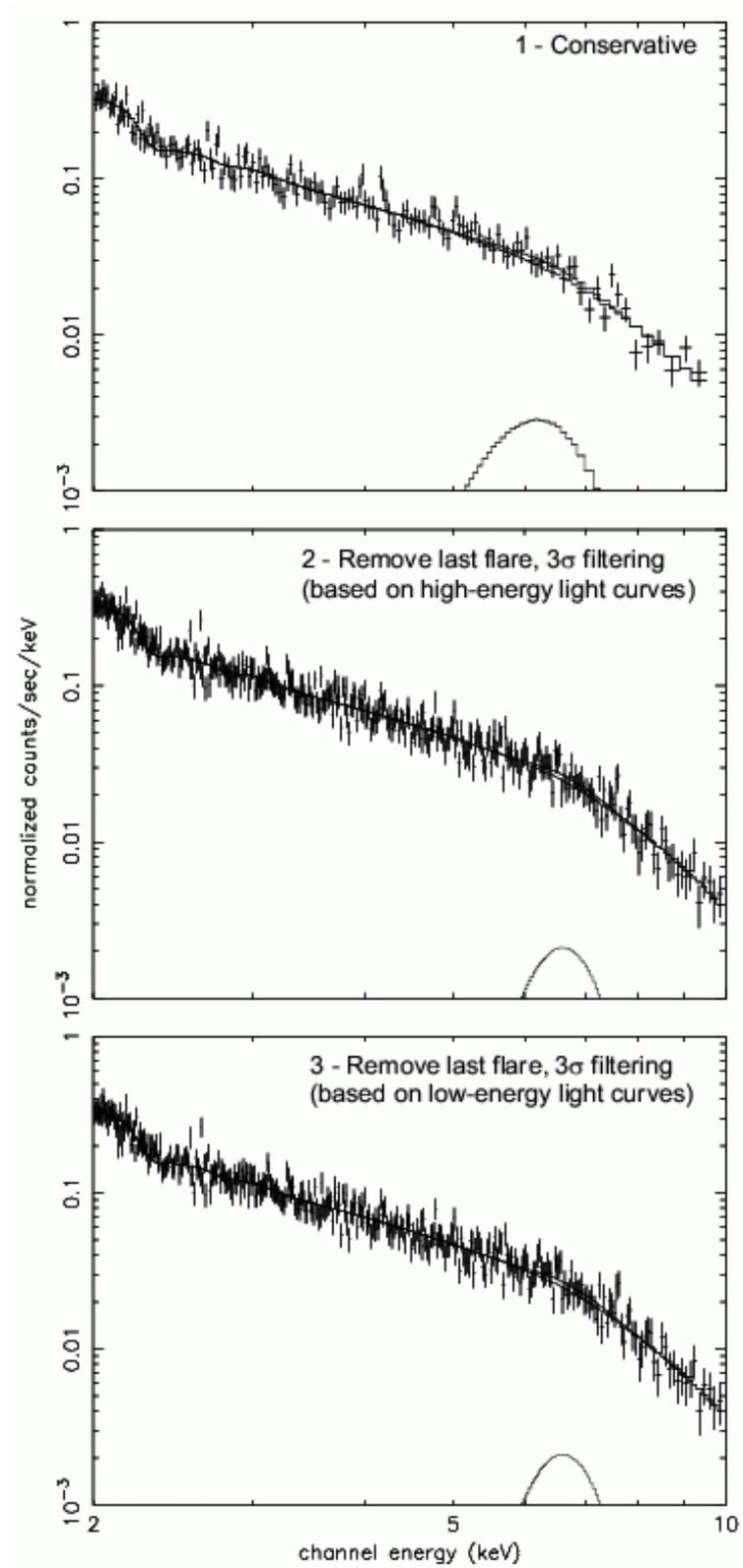}
\caption{Counts spectra for the pn camera, showing the model contributions from an absorbed power law, a Gaussian emission line, and their sum. Shown are the three GTI filtering methods used for the \XMM nuclear spectral analysis, as defined in Section 3.2.}
\label{xmm_gti_combined}
\end{center}
\end{figure*}

We also used the \XMM data to search for possible fluorescent line
emission from iron, and considered the statistical significance of its
detection for each GTI-filtering method described above. An absorbed
power law was fitted to the spectrum over an energy range of 2.0--5.0
keV and 7.5--10.0 keV, and the interpolated continuum subtracted from
the data in the energy range 5.0--7.5 keV. The continuum parameters
were consistent with those described above, and provided a good fit to
the data. The energy range was then extended to cover 5.0--7.5 keV,
and the continuum parameters frozen. A Gaussian Fe emission line with
a centroid allowed to vary between 6.0 and 6.9 keV and free line width
was then added to the model, and the significance of its addition
assessed with an $F$-test. This spectral analysis was performed for
the MOS1+MOS2 cameras jointly, the pn camera only, and finally all
three jointly. A summary of the significances is given in
Table~\ref{fe_sig_tab}, the best-fitting parameters are shown in
Table~\ref{fe_results_tab}, and counts spectra for the pn camera are
presented in Figure~\ref{xmm_gti_combined}. 

Our results show that a continuum-only model provides an adequate fit
to the spectra, and that any Fe K$\alpha$ detection is  entative, but
not highly significant, even when an unconservative GTI-filtering
method is used. Figure~\ref{xmm_gti_combined}, in particular,
illustrates that different methods of extracting light curves for
GTI filtering purposes, i.e., using high-energy or low-energy light curves,
has no effect on our conclusions. We also note that the only evidence
for the existence of the Fe K$\alpha$ line is in the data from the pn
camera; however, this is perhaps not surprising, as the effective area
of the pn camera is still somewhat higher than that of the MOS1+MOS2
cameras combined.

\begin{figure*}
\begin{center}
\includegraphics[width=8cm]{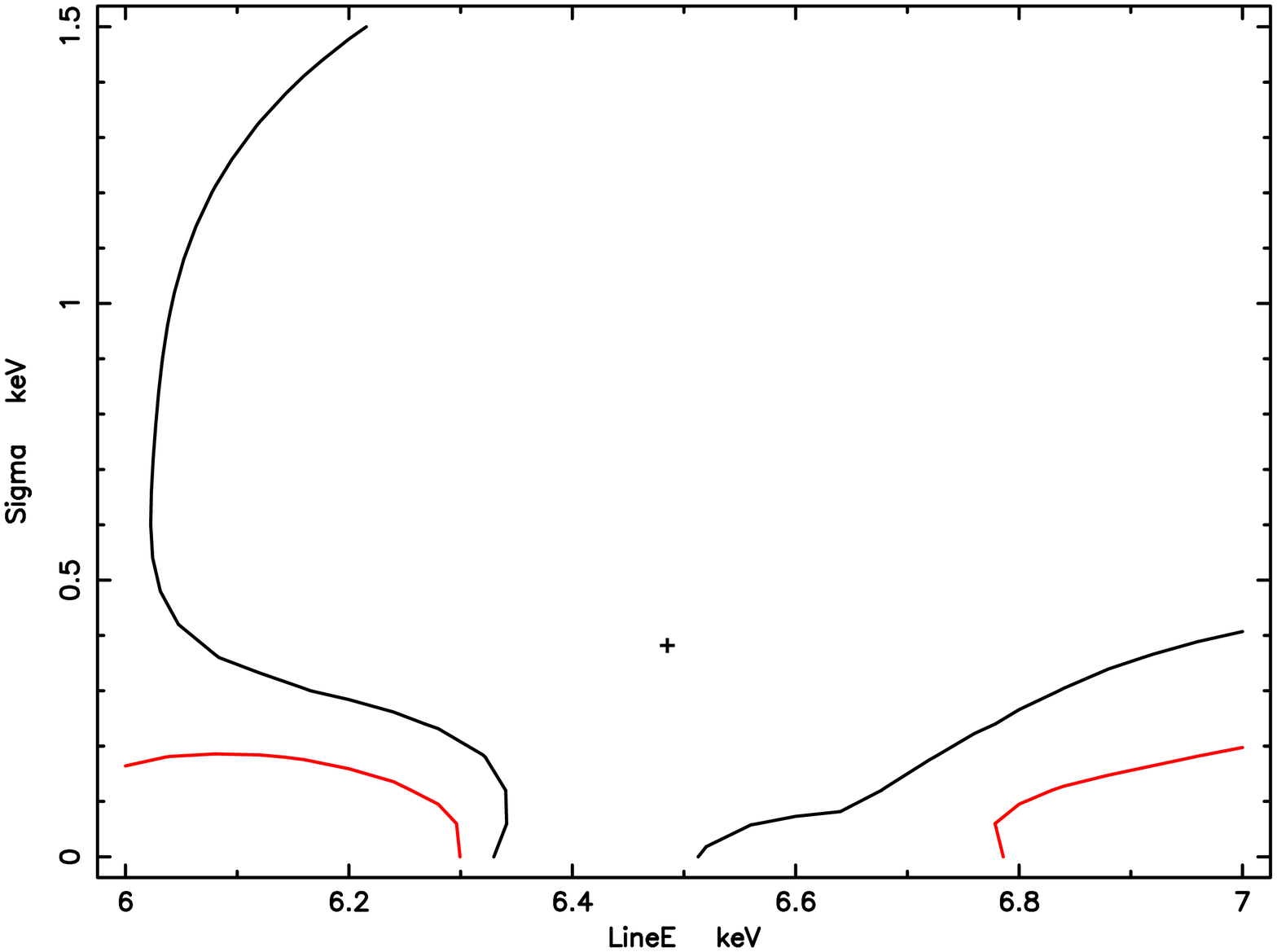}
\includegraphics[width=8cm]{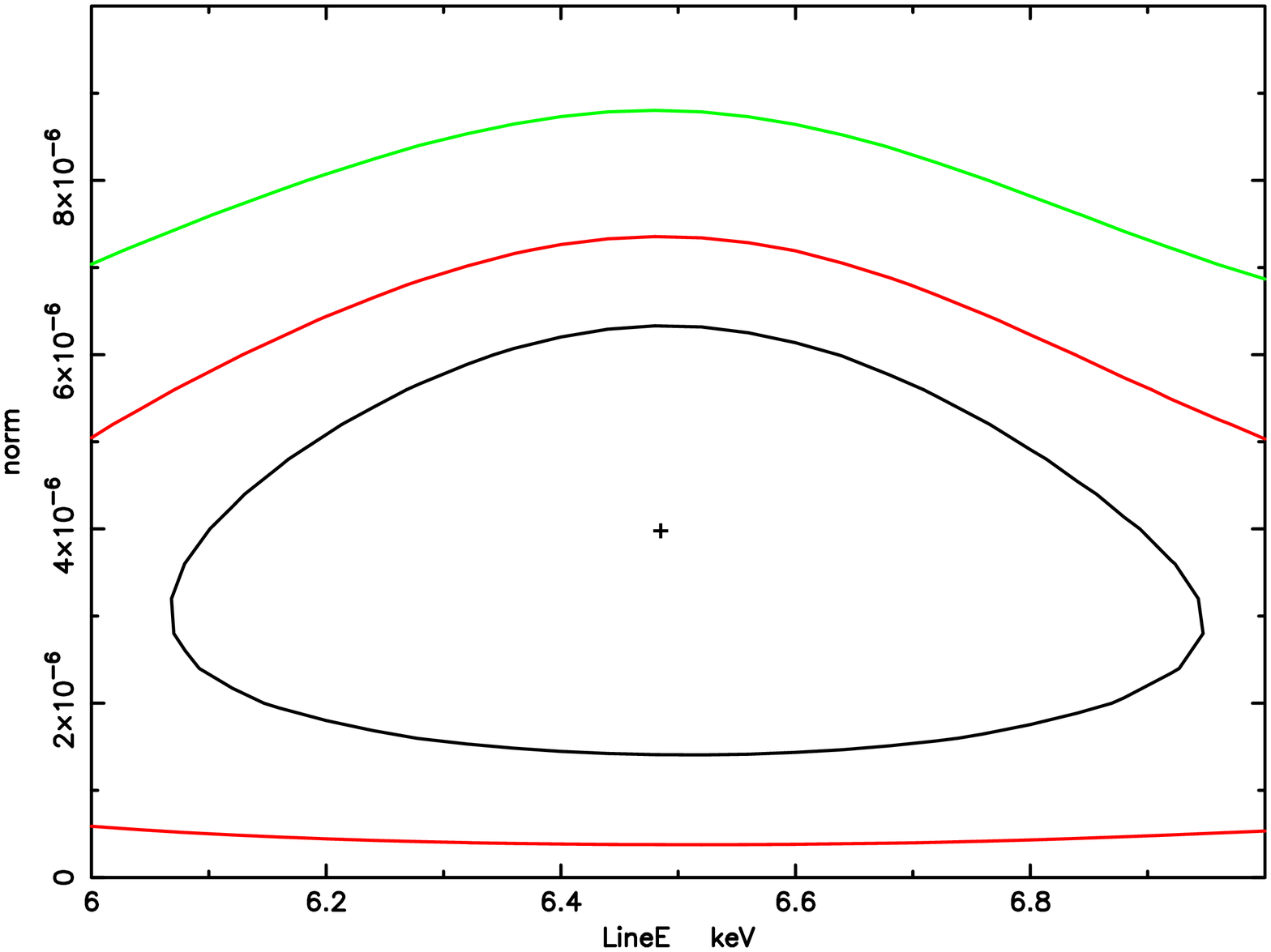}
\caption{Left: Energy and line width confidence contours of the Fe K$\alpha$ line for the pn spectrum using the 3$\sigma$ GTI-filtering method based on the low-energy light curve. Right: Energy and normalization confidence contours of the Fe K$\alpha$ line, after freezing the line width at 0.4 keV and reperforming the spectral fitting. This highlights the poor constraints that can be placed on the Fe K$\alpha$ line-parameters, even though an unconservative GTI-filtering method was used.}\label{pn_ref2_fe_contour}
\end{center}
\end{figure*}

For a more direct comparison with the results of \cite{gli04}, we
chose to investigate further the properties of the Fe K$\alpha$ line
measured in the spectrum of the pn camera. We chose the
3$\sigma$-filtering criterion based on the low-energy (0.4--10 keV) light
curves described in Section \ref{xmm}, which most closely matches the
GTI-filtering method used by \cite{gli04}. We created confidence
contours of the line energy and line width, as shown in
Figure~\ref{pn_ref2_fe_contour}. This test shows that the line
parameters are very poorly constrained. Even when the line width was
frozen at 0.4 keV and the fitting reperformed, the confidence contours
of the line energy and normalization (Fig.~\ref{pn_ref2_fe_contour})
show that the line parameters are still highly uncertain.

\subsubsection{Source variability}

a) X-ray

For the \XMM data, we attempted to confirm the intra-observation
variability claimed by \cite{gli04} by creating X-ray light curves for
each of the MOS1, MOS2, and pn cameras, using a source-centred
circular extraction region of radius 35 arcsec. Background subtraction
was applied using a source-centred annular extraction region of inner
radius 140 arcsec and outer radius 280 arcsec for the MOS cameras, and
an off-source rectangle for the pn camera. Only events with FLAG=0,
PATTERN $\leq 4$ (MOS) and $\leq$ 12 (pn), and an energy 0.8--10 keV
were included in our analysis. In this instance, the good time
intervals were chosen to be the same as those used by \cite{gli04}.

The \XMM data were binned to a variety of timescales, and a $\chi^{2}$
analysis used to test the null hypothesis. Statistically significant
variability was only found with a long bin time. With bins of 3 ks, we
found that the variability found in the data from the pn camera was
significant ($\chi^2 = 22.35$ for 13 dof; null hypothesis probability
= 95.0 per cent). Further, a co-added light curve for all three
cameras was produced and binned to 3 ks. We found the variability to
be significant ($\chi^2 = 26.07$ for 12 dof; null hypothesis
probability = 99.0 per cent). We do, however, caution that times
corresponding to periods of high background are present in this light
curve. While the background level is significantly lower than the
source count rate, we cannot be certain of its impact on the extracted
light curves. Background-subtracted results for the individual \XMM
cameras, as well as the sum of their data, are given in
Table~\ref{chandra_xmm_vartab}.

For the \Ch data, we attempted to search for variability on 3-ks
timescales by extracting an ACIS-S light curve from a source-centred
circle of radius 1.23 arcsec, with an off-source background extraction
region that matched the choice of background for the radial profiling
(see Section 3.4). Unlike the \XMM observations, no significant
variability was found on 3-ks timescales ($\chi^2 = 13.37$ for 16
dof), although this might be due to the limited sensitivity of \Ch
compared with {\it{XMM-Newton}}.  We note that fitting a line of
gradient of 8.31 $\times 10^{-4}$ cts s$^{-2}$ to the data is
significantly preferred ($\Delta\chi^2 = 7.77$ for one additional
parameter) over a uniform source brightness. Background-subtracted
results for \Ch are given in Table~\ref{chandra_xmm_vartab}. Finally,
we constructed a light curve of the hardness ratio, $H-S/H+S$, where
$S$ is the 0.7--1.5 keV count rate and $H$ the 3--7 keV count rate,
binned to 3 ks. The variability of the hardness ratio was not
significant ($\chi^2 = 13.62$ for 16 dof).

Finally, in Table~\ref{longterm_vartab}, we tabulate the long-term
2--10 keV power-law luminosity and photon-index history of
NGC~6251. The 2--10 keV power law luminosity has varied by a factor of
$\sim$5 during 1991--2003, while not all of the photon-index values
are mutually consistent. The \Ch photon index of $1.67 \pm 0.06$ is
somewhat flatter than that measured with {\it{XMM-Newton}}, but it is entirely
plausible that variability has occurred.\newline

\noindent b) Radio

\label{vlbi}
High-resolution VLBI observations of NGC~6251 (e.g.,
\citealt{jonetal86}) show a compact, self-absorbed core with a highly
inverted spectrum and a parsec-scale jet. Emission from the core does
not greatly dominate over that from the jet in the VLBI observations,
even at the highest available frequency of 10.7 GHz, where it is most
prominent. We used VLBI observations of NGC~6251 (D. Jones, private
communication) to estimate the flux densities of the core and jet at
several epochs and frequencies. The results are given in
Table~\ref{vlbi_vartab}. This shows that variability approaching a
factor of 2 has occurred on timescales of years.\newline

\noindent c) Optical

{\it Hubble Space Telescope} ({\it HST}) images of the central regions
of NGC~6251 have shown an unresolved optical core, surrounded by a
dusty disk (\citealt{fer99}). Optical core flux densities were
measured by \cite{har99} and \cite{chi03} from archival {\it HST}
observations. There now exist {\it HST} data from several epochs that
we can use to search for any optical and UV variability. We performed
photometry on multi-epoch data from the WFPC2 instrument (F555W and
F814W filters) and from the FOC instrument (F342W and F410M
filters). Following \cite{har99}, the source flux was determined using
a circular extraction region of radius 5 pixels (0.2 arcsec), using
background from an annular region of inner radius 5 pixels and outer
radius 10 pixels. As noted by \cite{chi03}, the largest source of
systematic error in determining the flux from the WFPC2 data is the
choice of background, as a result of the dusty disk surrounding the
nucleus. Unresolved nuclear emission in the FOC data dominates by a
large factor over extended emission, so the systematic errors may not
be as high. The measured count rates were converted to flux densities
at the central frequency of each filter, using the {\sc{SYNPHOT}}
photometry package and assuming a power law with $\alpha = 0.8$.

The reddening to NGC~6251 is the sum of Galactic and possibly
intrinsic components. The Galactic value was calculated using a
standard relationship between the neutral hydrogen column density and
the $B - V$ colour excess (\citealt{bur78}). We estimated the Galactic
$B - V$ colour excess to be 0.061 mag, in fair agreement with the
value calculated using the dust maps of \cite{sfd98} (0.087 mag). To
convert between the $B - V$ colour excess and the extinction, we used
the extinction curve of \cite{sch77}. As an example, this gives rise
to a B-band extinction, $A_B = 0.25$ mag. In addition, {\it{if}} the
hydrogen column that obscures the radio core (\citealt{wer02}) also
covers the optical core, we estimate a {\it{further}} extinction by
assuming the same extinction ratio as our Galaxy. 

The measured core flux densities for the multi-epoch {\it HST} data
are tabulated in Table~\ref{hst_vartab}, assuming both Galactic and
intrinsic absorption. This shows that the flux densities measured with
the WFPC2 instrument have varied by $< 10$ per cent over a period of
$\sim 1$ year. However, data from the FOC instrument are separated by
$\sim 5$ years, and show variability by a factor $\sim 3$.

\subsection{Jet}

X-ray emission from the jet of NGC~6251 comes from three regions: the
`inner jet', extending from about 1.5 arcsec to 27 arcsec from the
nucleus, which corresponds to the {\it ROSAT} HRI detection of
\cite{har99}; and two regions of the outer jet, which
we call regions 1 and 2, corresponding to `knots' 1 and 2 of
\cite{ker03}, and extending respectively between about 200--265 arcsec
and 330--410 arcsec from the core. The new {\it Chandra}/ACIS-S
observation clearly detects the inner jet and region 1
(Fig. ~\ref{wholej}), but region 2 falls off the CCD. The {\it
Chandra}/ACIS-I data give a detection of all three regions (Fig. 1 of
\citealt{ker03}), though the inner jet is only weakly detected because
of the positioning of core and jet in the area of the central chip
gap. Between the end of the inner jet and the start of region 1,
(i.e., between 30 and 200 arcsec from the core) there is a
3--4$\sigma$ excess of counts over the local background in the ACIS-S
data (about 50 0.5--5 keV counts), suggesting that there is some X-ray
emission associated with this region of the radio jet (the `middle
jet'). The data are not sensitive enough to allow us to associate this
detection with a particular component of the radio emission. In the
\XMM data both components of the outer jet (region 1, discussed
by \citealt{sam04}, and region 2) are detected, although the spatial
resolution of \XMM means that the inner jet is confused with the
core. For the \XMM analysis of the jet and extended emission, we used
the `conservative' GTI filtered data described in Section 3.2.1b.

\begin{figure*}
\begin{center}
\includegraphics[width=10cm]{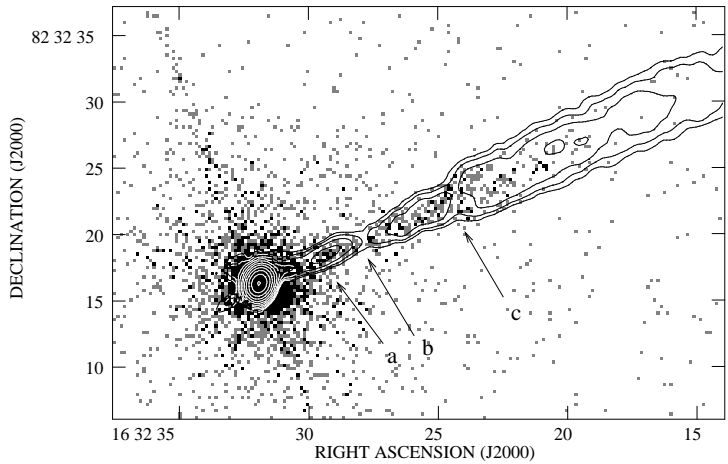}
\caption{The inner jet of NGC 6251. The greyscale shows {\it Chandra}
  X-ray counts between 0.5 and 5 keV, binned in 0.246-arcsec pixels;
  black is 2 counts pixel$^{-1}$. Superposed are contours from a radio
  map made from the A- and B-configuration VLA data at 1.6 GHz, with a
  resolution of $1.36 \times 1.09$ arcsec, at $(1, 2, 4\dots) \times
  0.4$ mJy beam$^{-1}$. The arrows show (a) a bright knot at 5 arcsec
  from the nucleus visible in both radio and X-ray, (b) a dip in the
  surface brightness at both wavebands, and (c) a bright region of the
  X-ray jet coincident with a dip in the radio surface brightness at
  15 arcsec from the nucleus.}
\label{innj}
\end{center}
\end{figure*}

\begin{figure*}
\begin{center}
\includegraphics[width=10cm]{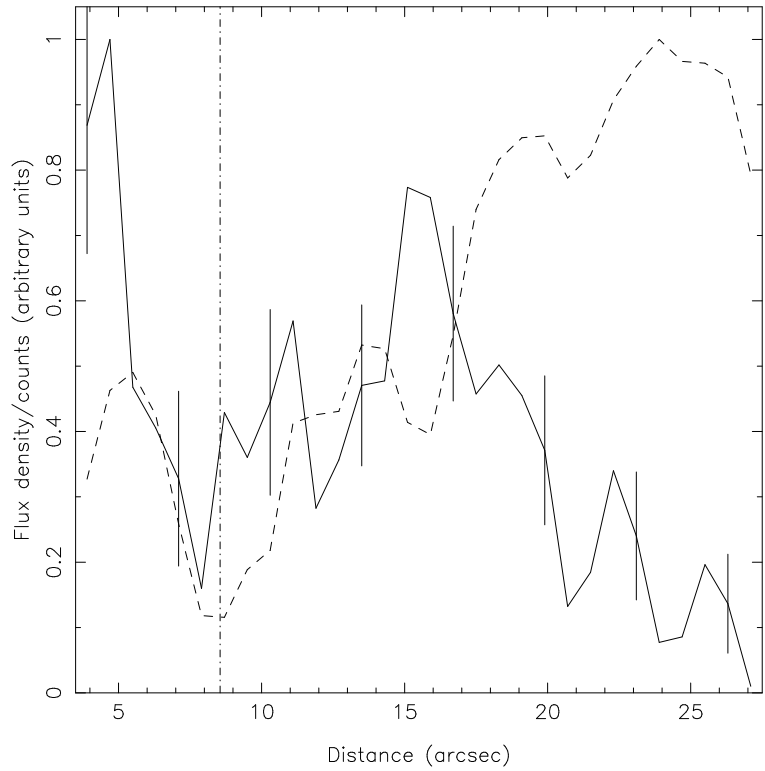}
\caption{A profile along the inner jet of 6251 in radio (dashed line)
  and X-ray (solid line). Data are taken from a rectangular strip 5.7
  arcsec wide, with adjacent background subtraction in the case of the
  X-ray data, and binned in 0.8-arcsec bins. The vertical dot-dashed
  line shows the position of the dip in the radio surface brightness
  ('b' on Fig. \ref{innj}). The vertical lines on the X-ray points
  indicate the Poisson errors.}
\label{innj-profile}
\end{center}
\end{figure*}

The relationship between the radio jet and the inner X-ray jet is
shown in Figure~\ref{innj}. The inner jet, which contains $380 \pm 20$
0.4--7.0 keV {\it Chandra}/ACIS-S counts, appears similar to the radio
emission; for example, a bright knot (`a') at 5 arcsec from the nucleus is
present in both radio and X-ray images, while there is a dip in both
radio and X-ray surface brightness at around 9 arcsec (`b'). At around 15
arcsec from the nucleus (`c'), the X-ray jet brightens while the surface
brightness of the radio jet drops, and there are hints that the X-ray
jet, which is transversely resolved with {\it{Chandra}}, may be
edge-brightened at this point; thereafter the flux in X-rays drops off
rapidly with distance along the jet (Fig. \ref{innj-profile}). A
spectral fit to the whole jet region, using an adjacent background of
the same shape and at the same distance from the nucleus, is well
fitted ($\chi^2 = 7.6$ for 15 dof) with a single power-law model
absorbed with the Galactic column density; the photon index is
$2.29_{-0.13}^{+0.14}$ ($1\sigma$ for one interesting parameter) and
the 1-keV flux density of this region of the jet is $10\pm 1$ nJy. A
thermal ({\sc{mekal}}) model with moderate (0.3 solar) abundance is a much
worse fit, with $\chi^2 = 28.5$ for 15 dof. There is no evidence for
differences in the power-law index as a function of position in the
jet; if we divide it into inner and outer halves containing
approximately equal numbers of counts, the fitted power-law indices
for the two halves are $2.6 \pm 0.3$ and $2.2 \pm 0.3$ respectively.

The two outer regions of the X-ray jet correspond to brighter features
in the low-frequency radio image, and the X-ray structure of region 1,
at least, is well matched to the radio (Fig.\ \ref{wholej}). There is
no sign of knotty or compact structure in either the radio or X-ray;
the compact feature appearing at about 230 arcsec from the nucleus in
Figure 5 of \cite{jon94} does not appear in our reduction of their VLA
dataset, and we suspect it to have been an imaging artefact. We
extracted spectra for region 1 from the two \Ch datasets, using
rectangular regions with adjacent off-source background, and also from
the three \XMM cameras, using a similar but somewhat broader region to
account for the larger \XMM PSF. A fit to the ACIS-S dataset alone
gives a good fit to a power-law model with Galactic absorption and
photon index $1.96_{-0.21}^{+0.23}$, but a joint fit to all five
datasets in the 0.4--7 keV energy range gives a good fit ($\chi^2 =
30.4$ for 26 dof) with lower, though consistent, power-law index
($1.68\pm 0.13$) and a 1-keV flux density of $4.7 \pm 0.4$ nJy. We
tested the significance of the detection for region 2 by comparing
source and background counts for the three \XMM cameras. We measured
205 background-subtracted 0.4--7.0 keV counts in the source region,
and 1686 counts in the background (when scaled to the size of the
source region), giving a $> 3\sigma$ level detection (the detection is
also significant for each separate camera). The best-fitting power-law
spectrum ($\chi^2 = 8.2$ for 7 dof) fitted to the three \XMM cameras
in region 2, again using a rectangular extraction region, has a photon
index of $1.7^{+0.8}_{-0.7}$, with a 1-keV flux density of
$3.4^{+1.0}_{-1.1}$ nJy. In both these cases the data can be fitted
equally well with thermal models with moderate abundance, but high
temperatures are required (5 keV in the case of region 1, even higher
in region 2).

\subsection{Extended emission}
 
\begin{figure*}
\begin{center}
\includegraphics[width=10cm]{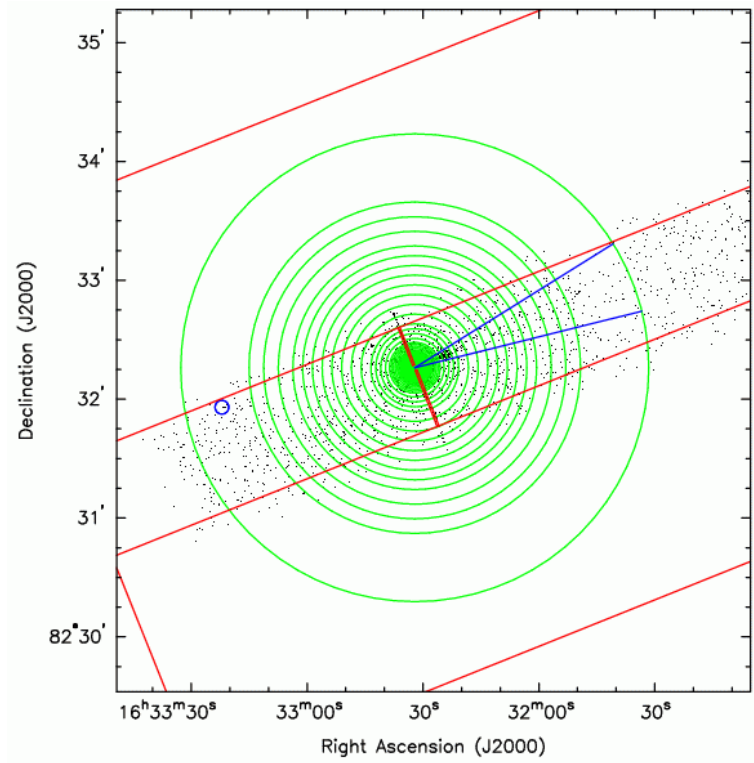}
\caption{Regions used in the {\it Chandra}/ACIS-S radial profiling analysis. Concentric circles of radii 0.98--118 arcsec (2--240 pixels) are shown. Background was extracted from the last annulus (170--240 pixels). The regions excluded from the analysis were the two large outer rectangles, the two small rectangles (masking the frame transfer streak), a small circle (masking a point source), and a pie slice between position angles 284 and 302 degrees (avoiding contamination from jet emission).}
\label{acis_evt2_0.5-5.regions}
\end{center}
\end{figure*}

To search for any extended emission on 10--100 kpc scales, a radial
profile was extracted using data from the {\it Chandra}/ACIS S3 chip
in the energy range 0.5--5.0 keV. The profile was centred on the
nucleus, and excluded one detected background point source, as well as
jet emission, which was excluded by ignoring a pie slice between the
position angles 284 and 302 degrees. The radial profile extended from
0.98 arcsec (2 pixels) out to 84 arcsec (170 pixels), with background
taken from an annulus of inner radius 84 arcsec and outer radius 118
arcsec (240 pixels) (see Figure~\ref{acis_evt2_0.5-5.regions}). A PSF
was modelled using ChaRT and MARX, and was convolved with a Gaussian
of r.m.s. width 0.35 arcsec in order to provide a good match with the
radial profile data close to the core.

\begin{figure*}
\begin{center}
\includegraphics[width=10cm]{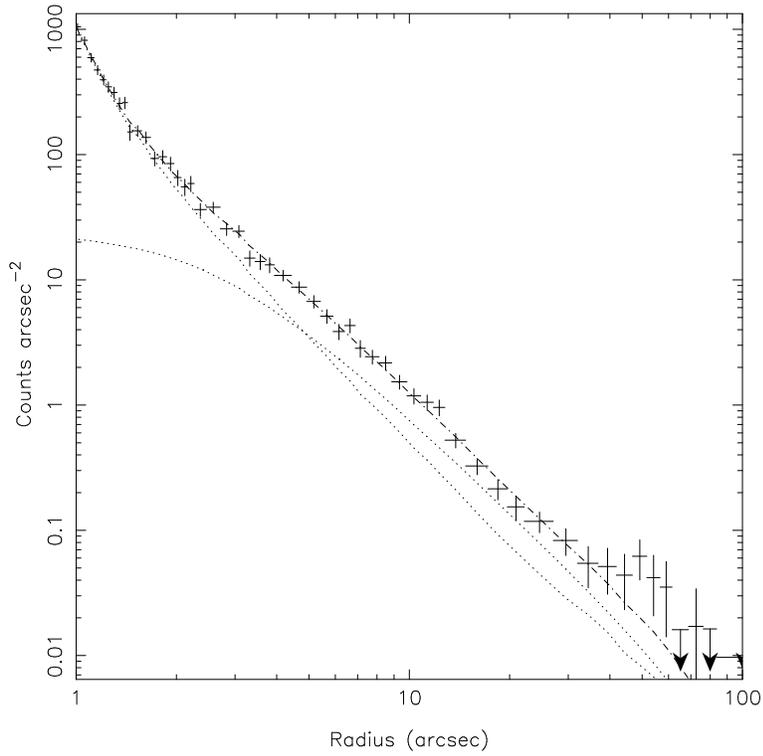}
\caption{Background-subtracted 0.5--5 keV radial surface brightness profile for the {\it Chandra}/ACIS-S data. The best-fitting model is a composite of a point-like component and a $\beta$-model, with the best-fitting $\beta$-model parameters described in the text.}
\label{acis_evt2_0.5-5.radial.bestfit}
\end{center}
\end{figure*}

The radial profile analysis gave an unacceptable fit to
single-component models of either a point source or a $\beta$-model
(used to describe gas in hydrostatic equilibrium), but a combination
of the two gave an acceptable fit ($\chi^{2}_{\nu}$ = 0.79), as shown
in Figure~\ref{acis_evt2_0.5-5.radial.bestfit}. The best-fitting
parameters of the $\beta$-model were $\beta = 0.595^{+0.065}_{-0.045}$
and $\theta_c = 2.527^{+1.393}_{-1.251}$ (all errors are $1\sigma$ for
two interesting parameters, i.e., $\chi^2_{\rm min} + 2.3$). The FWHM
of the gas distribution is $\sim 2$ kpc, which is comparable to that
seen in other radio-loud ellipticals, such as 3C\,31 (1.7 kpc,
\citealt{har02a}). In our annular spectral extraction regions of radii
0.492 -- 1.23 arcsec and 0.492 -- 35 arcsec (see Section 3.2.1), we
would expect to see $94^{+123}_{-39}$ ($^{+220}_{-49}$) and $1499 \pm
315$ ($\pm 494$) thermal counts respectively, where uncertainties are
quoted as $1\sigma$ (unbracketed) and 90 per cent (bracketed), for two
interesting parameters.

\begin{figure*}
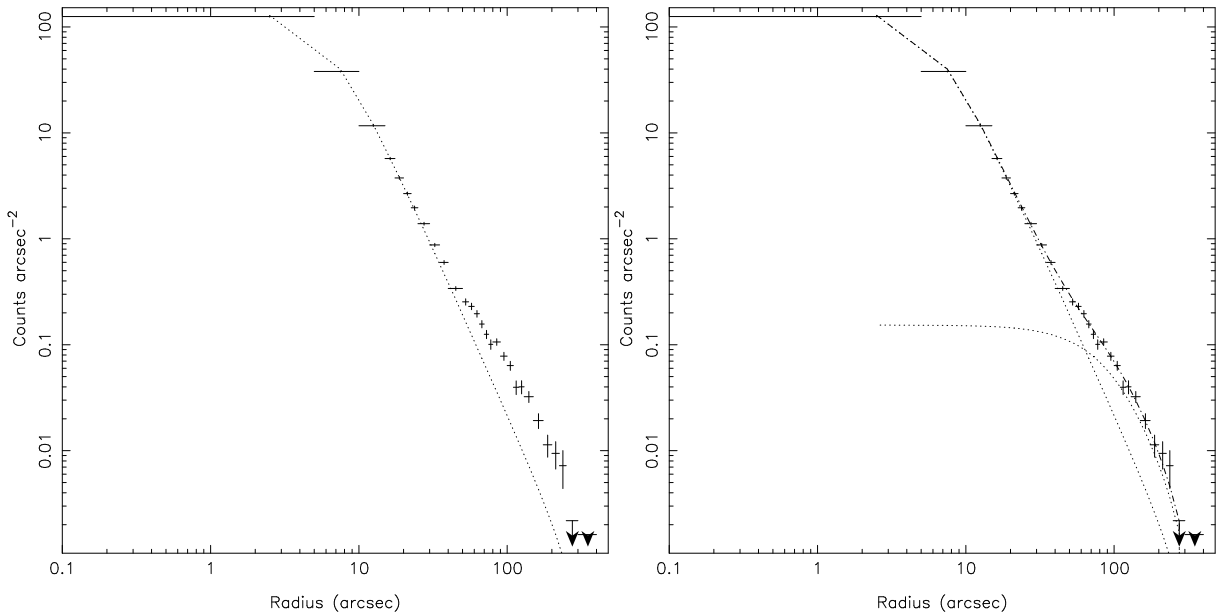

\begin{center}
\centering{\hbox{
\includegraphics[width=8cm]{f12a.eps}
\includegraphics[width=8cm]{f12b.eps}}}
\caption{Radial surface brightness profiles for the pn data in the
  0.3--7 keV energy band. Left: single point-source model. Right:
  convolved $\beta$-model plus point source, with the joint
  best-fitting $\beta$-model parameters given in the text.}
\label{prof}
\end{center}
\end{figure*}

The extended emission on larger scales, previously discussed by
\cite{bw93}, \cite{mac97a} and \cite{sam04}, is clearly seen in our
maps of the \XMM data (Fig.\ \ref{ext_overlays}). The emission is
roughly radially symmetric and centred on the radio source, though
there is an X-ray deficit to the SW of the nucleus (also seen by
\citealt{mac97a}) that does not appear to be related to any of the radio
structure. To characterize this emission, we extracted radial profiles
for the three \XMM cameras and for the {\it Chandra}/ACIS-I data. The
profiles were obtained in the energy range of 0.3--7 keV out to a
radius of 450 arcsec; the inner 2 arcsec of the \Ch profile was
excluded to avoid the chip gaps. We fitted single point-source models
and convolved point-source plus $\beta$-models to the \XMM profiles.
For all profiles, the point-source models were not an adequate fit to
the data, so that the inclusion of the $\beta$-model component is
needed. In all cases, the inclusion of a $\beta$-model of suitable
parameters leads to good values for the fit statistic (an improvement
from $\chi^2 = 132$ for 43 dof to $\chi^2 = 47.9$ for 42 dof for MOS1,
from $\chi^2 = 111$ for 46 dof to $\chi^2 = 46.7$ for 45 dof for MOS2,
and from $\chi^2 = 677$ for 28 dof to $\chi^2 = 26.5$ for 27 dof for
pn). We tested the significance of the improvement separately for each
camera using Monte Carlo simulations (as in \citealt{cro04a}), and in
each case we found a probability of $< 0.1$ per cent that the
improvement in $\chi^{2}$ given above could occur by chance.
Figure~\ref{prof} shows the pn profile with a point-source model and
the best-fitting point-source plus $\beta$-model. For the ACIS-I
dataset, we included a fixed small-scale $\beta$-model based on the
ACIS-S data in the fit, and were then able to obtain adequate fits
with an additional large-scale $\beta$-model component. We find the
best-fitting parameters for a joint fit to the ACIS-I, MOS1, MOS2 and
pn profiles to be $\beta = 0.90_{-0.25}^{+0.6}$, $r_c =
116_{-38}^{+64}$ arcsec (errors are $1\sigma$ for 2 interesting
parameters), which are consistent with the best-fitting models found
by \cite{bw93} for the {\it ROSAT} data. The individual datasets all
give best-fitting parameters that are consistent at the joint
$1\sigma$ level. The joint fit is good ($\chi^2 = 152$ for 134
degrees of freedom). The residuals are dominated by systematic
uncertainties in the modelling of the \XMM PSF and in the background
subtraction at large radii.

We extracted spectra from the three \XMM cameras to study the extended
emission identified by the radial profile analysis, using the
`conservative` GTI-filtering described in Section 3.2.1b. We used an
extraction annulus between 60 and 300 arcsec and used the double
subtraction technique described in Section 2.2 to obtain appropriate
background spectra. The radial profile analysis shows that there is a
significant contribution from the bright central point source at these
radii. We included this contribution in the spectral fits as a fixed
component with the parameters of our best-fitting core spectrum (an
{\it apec} model of $kT=0.60$ keV and $Z=0.35$ solar, and a power-law
with $\Gamma=1.85$), scaling the normalizations by the fraction of
encircled energy in the extraction region (8.3 per cent). In addition
to this fixed component, we fitted a second {\sc apec} model for the
extended emission. We found a best-fitting temperature of
$1.7^{+0.4}_{-0.3}$ keV and abundance of $0.4^{+0.4}_{-0.2}$ solar
(both consistent with the results of
\citealt{sam04}), with $\chi^2$ of 292 for 305 degrees of freedom.

For all three \XMM cameras, the total counts in the $\beta$ model
between radii of 60 and 300 arcsec are consistent with the counts in
the thermal component of the best-fitting spectrum described above.
The total bolometric luminosity for the group atmosphere, determined
by integrating the $\beta$-model, is found to be $(7\pm1) \times
10^{41}$ ergs s$^{-1}$.

\section{Interpretation}

\subsection{Core}

The nuclear continuum spectrum of NGC~6251 detected with \Ch and \XMM
is well described by an absorbed power-law model. This result is
consistent with two distinct models for the nuclear X-ray emission
from radio galaxies. The first is that the non-thermal X-radiation is
dominated by emission from an accretion flow, via inverse-Compton
upscattering of a population of lower-energy photons to X-ray
energies. Such a model would be supported by the detection of
broadened fluorescent Fe K$\alpha$ emission of high equivalent
width. However, our analysis of the \XMM observation first presented
by \cite{gli04} has shown that the detection of the Fe K$\alpha$ line
is not {\it{highly}} significant (a continuum-only model provides an
adequate fit to the spectrum), leading us to disfavour such a
model. The line is not detected with {\it{Chandra}}, although this is
possibly related to the relatively poor photon statistics of the ACIS
camera at 6.4 keV. Further, the poorly constrained parameters using
the \XMM data mean that it is impossible to constrain its origin, much
less to argue that it originates in the inner regions of an accretion
disk.

A second possible interpretation of the data (\citealt{bw93,chi03}) is
that the X-ray emission arises primarily from the subparsec radio VLBI
jet. To investigate this model further, we calculated the flux density
ratio of the 1-keV power-law X-ray emission measured with \Ch and the
parsec-scale radio jet seen in 5-GHz VLBI observations
(\citealt{jonetal86}). We performed a similar calculation for the
inner (kiloparsec-scale) jet, using our {\it Chandra}/ACIS-S
observation and 8-GHz VLA observations. As shown in in
Table~\ref{fluxratios_tab}, the slopes and flux-density ratios of the
X-ray and radio components of the parsec- and kiloparsec-scale
emissions are approximately similar. This similarity also occurs with
the nearby radio galaxy Centaurus A, where the soft nuclear X-ray
component is interpreted as having a jet origin
(\citealt{evans04}). The similar observed variability on timescales of
years of the radio, optical, and X-ray data might also suggest a
common origin for the nuclear emission, although we note that the data
are non-contemporaneous, so no firm conclusions can be drawn.

Previous studies of the X-ray and radio flux properties of the active
nuclei of low-power radio galaxies (\citealt{fab84,wor94,can99,har99})
have shown a correlation between the luminosities of soft, unresolved
X-ray emission and 5-GHz core radio emission. NGC~6251 lies near the
trendline established for the large B2 sample of radio galaxy cores
(\citealt{can99}). This is consistent with the hypothesis that the
observed X-ray emission is associated with the parsec-scale VLBI
jet. This is not the only possible interpretation. For example,
\cite{mer03} find evidence for a correlation between the X-ray and
radio luminosities, together with the black hole mass, of a sample of
100 AGN, and relate the X-ray emission to an accretion flow. However,
the majority of their target objects are not radio galaxies, and so
may not possess relativistic parsec-scale jets.

\begin{figure*}
\begin{center}
\includegraphics[width=10cm]{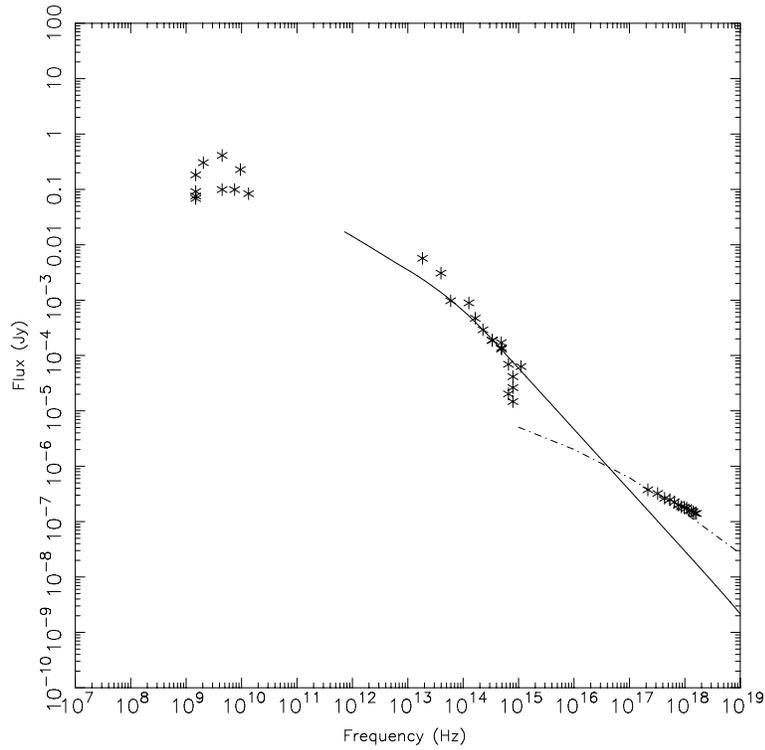}
\caption{Radio-to-X-ray spectrum of NGC~6251 in the source frame, with
  model synchrotron and SSC fits to the data assuming a spherical
  geometry with emitting radius $5 \times 10^{-4}$ pc. The solid line
  discussed shows the best-fitting synchrotron model, and the
  dot-dashed line shows the SSC emission expected. The electron
  spectrum is assumed to extend from $\gamma_{\rm min} \sim 20$ to
  $\gamma_{\rm max} \sim 4.5 \times 10^{5}$ with a number spectral index
  of $p = 2.4$, breaking by 1 at $\gamma \sim 600$. The model
  parameters are representative only and are poorly constrained due to
  the historical variability of the source and non-contemporaneous
  nature of the data. The model synchrotron emission shown is
  truncated at a frequency $10^{12}$ Hz, as self-absorption becomes
  important below this frequency.}
\label{8jb}
\end{center}
\end{figure*}

In order to place further constraints on the X-ray emission mechanism, we
constructed a spectral energy distribution of the nucleus of NGC~6251,
using observations from VLBI (D. Jones, private communication), ISO
(Birkinshaw et al., in preparation), {\it HST} (present work;
\citealt{har00,chi03}), \Ch (present work), \XMM (present work), {\it
BeppoSAX} (\citealt{gua03}), {\it ASCA} (\citealt{tur97}), {\it ROSAT}
(\citealt{bw93}), and {\it {CGRO}}/EGRET (\citealt{muk02}). {\it HST}
data were dereddened following the method described in the preceding
section. The SED is double-peaked, and is similar to
previously-determined SEDs for NGC~6251 (e.g., \citealt{chi03}), where
the authors argued that the SED is consistent with a blazar-type
model, with the X-rays being produced by synchrotron-self Compton
(SSC) emission. Figure~\ref{8jb} shows a plausible SSC model fit to
the radio-to-X-ray emission of NGC~6251, which has been more
extensively discussed elsewhere (\citealt{chi03,gua03,ghi04}).

It is possible, however, that in this jet-dominated scenario there
exists a `hidden' accretion-related component of luminosity of the order
$10^{41}$ ergs s$^{-1}$. Such a model would permit a fluorescent
Fe K$\alpha$ line, and indeed for NGC~6251 we speculate that the (weak)
detection of the line is evidence for this model. We shall return to
this point elsewhere (Evans et al., in preparation). It is further
possible that the relative strength of jet- and accretion-related
components varies over time: as noted by \cite{gua03}, states of high
nuclear luminosity might correspond to periods when emission from a
relativistic jet dominates the observed X-ray flux, whereas during
lower-luminosity periods, emission from an accretion flow might be
most prominent. This is supported by the fact that the most
significant detection of a fluorescent Fe K$\alpha$ line, with ASCA in
1994 (\citealt{tur97,gua03}), occurred when the continuum flux was at
its lowest historical level (see Table~\ref{longterm_vartab}).

\subsection{Jet}

\begin{figure*}
\begin{center}
\includegraphics[width=10cm]{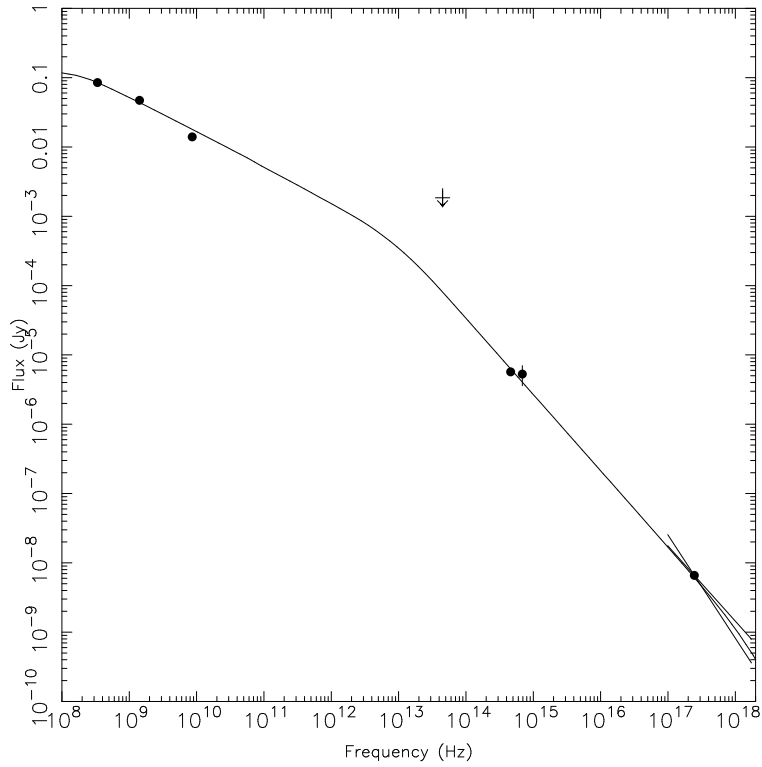}
\caption{The radio to X-ray spectrum of the inner NGC~6251 jet.
  Plotted are data points from the radio, an infra-red upper limit
  from ISO (D. Tansley, private communication), the optical
  measurements (Werner 2002) and the X-ray data from \Ch (with 1-keV
  flux density and spectral errors). The solid line is the fitted
  synchrotron spectrum described in the text.}
\label{spectrum-ij}
\end{center}
\end{figure*}

The natural interpretation of the inner jet is as a synchrotron jet
similar to those seen in other FRI-type radio galaxies
(e.g. \citealt{har01}). The steep X-ray spectrum is hard to reconcile
with any other emission mechanism and, as seen in
Table~\ref{jetxrflux}, the X-ray-to-radio flux ratio is similar to
those seen in the resolved jets of other low-power radio galaxies
where the jet emission is interpreted as synchrotron in
nature. Additional support for this model comes from the detection of
an apparent optical counterpart of the jet (\citealt{wer02}). The
optical jet is detected between about 8 and 20 arcsec from the
nucleus, and a simple synchrotron model with a broken power-law
electron energy spectrum can be fitted to the radio, optical and X-ray
data in this region without difficulty, with the low-frequency
spectral index $\alpha$ being 0.5 and the high-frequency value $1.1$
(Fig. \ref{spectrum-ij}). This model is similar to those fitted to
other radio, optical and X-ray jets.

For an equipartition magnetic field strength in this region (around 2
nT) the electron energy-loss timescale for an X-ray-emitting electron
is of the order of a few hundred years, while the light travel time to
the base of this region is $\sim 1.2 \times 10^4$ years, ignoring the
effects of projection and beaming (which may be considerable, as we
discuss below). The synchrotron origin of the X-ray emission from the
inner jet requires {\it in situ} particle acceleration in this
source. [\cite{ker03} come to a different conclusion, but their
calculation of the loss timescale is based on a magnetic field
strength which is inappropriate for the inner jet.] {\it In situ}
particle acceleration in turn requires an energy source. In the case
of the well-studied FRI jets in 3C\,31 it has been argued
(\citealt{har02a,lai04}) that the energy for the particle acceleration
that produces the observed X-ray jet comes from the bulk deceleration
of the jet flow, a process that plausibly happens for many other FRI
jet sources in which the jet-counterjet asymmetry decreases with
distance from the nucleus (e.g., \citealt{lai99}). If we believe that the
jet-counterjet asymmetry in the case of NGC~6251 (whose one-sided
radio jet is morphologically very similar to those seen in other FRI
sources) is due to relativistic beaming, then clearly the jet does not
decelerate to sub-relativistic speeds in the inner few kiloparsecs, as other
well-studied FRI jets in sources of similar luminosity apparently do,
but instead retains at least moderately relativistic speeds out to
hundreds of kpc. However, based on the radio sidedness evidence alone,
we cannot rule out a model in which some deceleration occurs in the
inner 13 kpc (projected) in which we see the inner X-ray jet, and in
which the deceleration provides the energy source for the required
particle acceleration.

The interpretation of the outer jet is less obvious. In region 1, the
synchrotron model that we fitted to the inner jet significantly
overpredicts the X-ray flux density; more importantly, the
best-fitting spectral index is too flat. The two-point spectral index
between 1.4 GHz and 1 keV is 0.95, while the best-fitting X-ray
spectral index is 0.71, which taken at face value rules out a one-zone
synchrotron model (the low-frequency radio spectral index is $\sim
0.5$). Spectral indices as steep as 1.09 are not ruled out at the 99
per cent confidence level, so we can devise a one-zone synchrotron
model in which the electron energy spectrum breaks immediately after
the radio region to a spectral index $\sim 1$ and then continues to
the X-ray. We do not regard this model as impossible, but it involves
ad hoc assumptions about the electron energy spectrum and is only
marginally consistent with the X-ray data. Region 1 is much larger
than the synchrotron loss scale for X-ray emission electrons in an
equipartition magnetic field strength, so we have no particular reason
to expect a one-zone synchrotron model to be applicable, but
nevertheless it is clear that region 1 is not well described by the
sort of model that has traditionally been fitted to the inner
synchrotron jets of FRIs, including NGC~6251.

\begin{figure*}
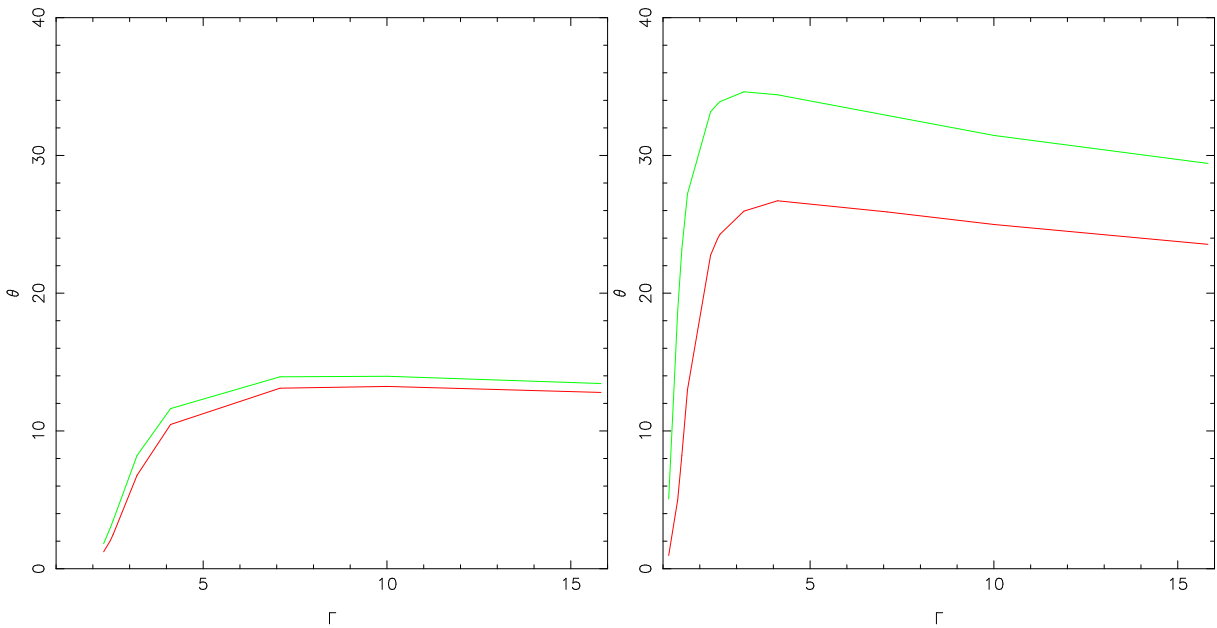

\begin{center}
\includegraphics[width=8cm]{f15a.eps}
\includegraphics[width=8cm]{f15b.eps}
\caption{Constraints on the bulk Lorentz factor and angle to the line
  of sight for region 1 (left) and 2 (right) of the outer jet. The
  allowed region of parameter space (given the $1\sigma$ range on
  X-ray flux density) lies between the two lines.}
\label{r1cons}
\label{r2cons}
\end{center}
\end{figure*}

Ignoring thermal models (which cannot explain the close relationship
between radio and X-ray emission) the alternative simple model for
region 1 is the one favoured by \cite{sam04}, in which the jet is
highly relativistic and the X-ray emission is caused by scattering of
the boosted microwave background radiation (e.g.,
\citealt{tav00}). This model naturally explains the flat spectral
index, which would be expected to be similar to the low-frequency
radio index (0.5--0.6). Such models have been used to describe the
jets in core-dominated Fanaroff-Riley type II (FRII) quasars, but in
general in FRIs we might expect the jets to be too slow to exhibit
significant effects, and the model also requires small values of the
angle $\theta$ of the jet to the line of sight. In NGC~6251, however,
we have already seen that the jet does not appear to decelerate to
sub-relativistic speeds on 10-kpc scales, so that it is possible that
the 100-kpc-scale jet could be produced by beamed inverse-Compton
emission. To investigate this, we calculated the expected
inverse-Compton flux density for a grid of bulk Lorentz factors and
angles to the line of sight $\theta$. We assumed a minimum electron
Lorentz factor of 10 and a maximum of $4
\times 10^5$, an electron power-law index of 2 (corresponding to a
spectral index $\alpha$ of 0.5) and Doppler boosting appropriate to a
continuous jet ($S \propto {\cal D}^{2+\alpha}$, where ${\cal D}$ is
the Doppler factor) and took account of the variation of the length of
the jet as a function of $\theta$. The anisotropic nature of the
inverse-Compton emission was accounted for using the code described by
\cite{har02b}, which implements the results of \cite{bru00}. The
result of our calculations was a grid of Lorentz factor and angle,
from which we were able to interpolate the range of allowed angles for
a given Lorentz factor, given the observed X-ray flux density. The
requirements for beaming parameters are quite extreme. A bulk Lorentz
factor as low as $\sim 2.5$ ($\beta \approx 0.92$) is possible, but
only if the jet is pointing almost directly along the line of sight,
which is clearly not the case. For more reasonable values of $\theta$,
the required bulk Lorentz factor is larger, and the model requires
$\theta < 15^\circ$ for $\Gamma < 16$ ($\beta < 0.998$) (Fig.\
\ref{r1cons}). These high bulk speeds would have to be maintained
across the well-resolved X-ray jet, $\sim 10$ kpc in width. Whereas
there is some evidence for a spine-sheath model involving relativistic
speeds on parsec scales in BL Lac objects (e.g. \citealt{chi00}), and
the spine speeds in such models are comparable to those required to
produce the observed level of X-rays through the beamed
inverse-Compton process, we emphasise that, if this were to persist to
Mpc scales (as required here) it would {\it not} produce the X-ray jet
structure seen in NGC~6251 -- instead, we would expect an X-ray jet
that was significantly narrower than the radio region.

Other constraints on the angle to the line of sight and jet speeds
come from analysis of the jet-counterjet ratio on the parsec and
kiloparsec scales: for example, the lower limit on the high-frequency
parsec-scale jet-counterjet ratio ($R$) determined by \cite{jon02},
$R>128$, would imply $\theta < 40^\circ$ if all the jet-counterjet
asymmetry were attributed to beaming. On larger scales, a weak
kiloparsec-scale counterjet is detected in some images, implying $R =
40$ (\citealt{per84}); if $\theta < 40^\circ$, this would imply $\beta
< 0.84$ on those scales. But in region 1 of the outer jet the
jet-counterjet ratio is $>200$, requiring either higher speeds or a
change in the angle to the line of sight. All of these arguments are
based on the assumption that the jet and counterjet are steady and
intrinsically symmetrical, which is dangerous, since some of the
properties of the large-scale jet are probably influenced by the
large-scale environment, as we shall argue later. Although the
one-sided parsec- and kiloparsec-scale jets provide some evidence for
relativistic speeds and angles to the line of sight $\la 40^\circ$,
they cannot tell us convincingly whether the constraints on the CMB
boosting model (Fig.\ \ref{r1cons}) are reasonable. Finally, it is
worth noting that NGC~6251 is already one of the largest known radio
galaxies, with a {\it projected} linear size close to 2 Mpc. It is
therefore {\it a priori} unlikely that the whole source structure is
close to the line of sight.

Region 2 provides few additional constraints. In this region, the
simple synchrotron model fitted to the inner jet gives a good
prediction of the observed 1-keV flux density (within the large
errors); the weak constraints on the X-ray spectrum mean that this
model is acceptable in that respect too. However, the large size of
the region means that only moderate amounts of beaming are required to
provide the observed X-ray in the boosted inverse-Compton model (Fig.\
\ref{r2cons}). Thus, both a synchrotron and inverse-Compton model are
possible without difficulty for this component. The inverse-Compton
model would require a change in jet speed and/or orientation, since
the constraints on beaming parameters are not consistent with those
determined for region 1. But we know that the jet does bend, in
projection, at around region 1, so this is not impossible. On the
other hand, the weakly detected middle jet (between 30 and 200 arcsec
from the core) does provide an important additional constraint. The
X-ray counts in the ACIS-S data for this region correspond to a 1-keV
flux density, assuming a similar spectrum to the other components,
around 1.4 nJy. But the large size of the middle jet region means that
it should be a relatively good source of beamed IC. With parameters
similar to those determined from region 1, we overpredict the observed
X-ray emission by a factor $\sim 5$. The inverse-Compton model can
only explain this if (a) the jet speeds up significantly between the
middle jet and region 1, or (b) there is a significant change in the
angle to the line of sight between the middle jet and region 1, in the
sense that the jet bends towards us at region 1, with all the bending
taking place perpendicular to the plane of the sky; for bulk Lorentz
factors around 5--7, the flux constraints require that the middle jet
have $\theta > 20\degr$. We regard both of these {\it ad hoc} assumptions as
unlikely.

We conclude that all possible models for the X-ray emission for region
1 of the jet are problematic to some extent, but that a synchrotron
model is more consistent with what is already known about FRI jets and
with the large-scale properties of the source. This then leaves us
with the question of why regions 1 and 2 are particularly good sites
for X-ray synchrotron emission and the corresponding {\it in situ} particle
acceleration. We return to the nature of these regions in the next section.

\subsection{Extended emission}

The detection of large and small-scale thermal components allows us to
investigate the long-standing question of jet confinement in NGC~6251.
We can estimate the minimum pressures in the jet components in the
standard way, choosing a conservatively low minimum electron energy
corresponding to $\gamma_{\rm min} = 10$, assuming a low-energy
electron energy index of 2 (which is consistent with the observed
low-frequency behaviour of the jet) and using the spectrum that fits
the radio, X-ray and optical data at higher energies. We assume no
significant contribution from relativistic protons, so that these are
true minimum-energy values. The remaining uncertainties are projection
and beaming effects. Beaming {\it in general} makes the jet appear
brighter than it really is, and so causes us to overestimate the
minimum pressure. Projection, which goes along with beaming, causes
us to underestimate the true size of components, so again we
overestimate their pressure: but it also means that they are further
out in the thermal atmosphere than we expect.

\begin{figure*}
\begin{center}
\includegraphics[width=10cm]{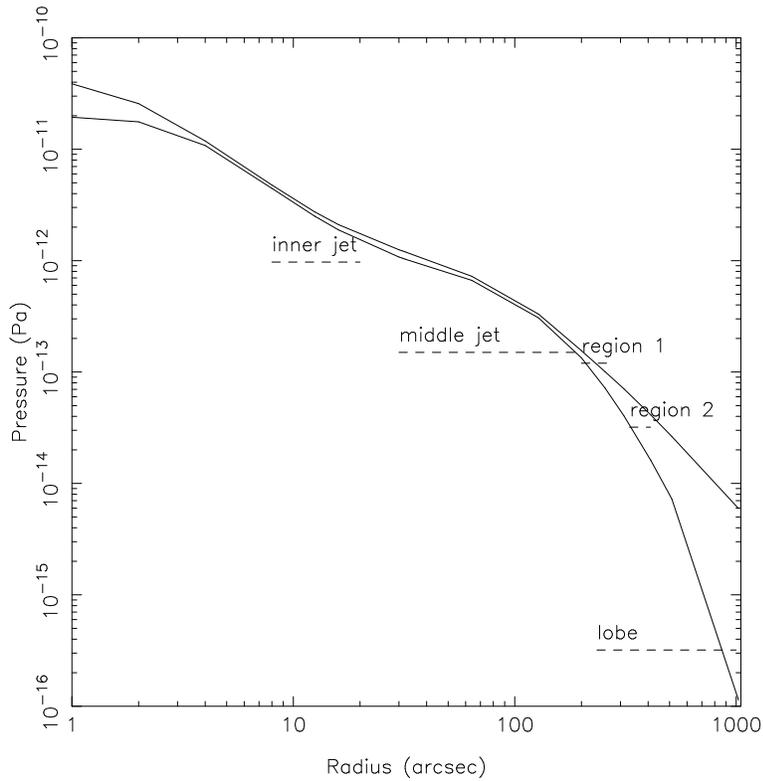}
\caption{The thermal pressure (solid lines) and minimum radio pressure
(dashed lines) in the radio components of NGC~6251 assuming no beaming
and a jet in the plane of the sky (see Table \ref{pressures}). The
solid lines show the total pressure in the thermal component, with the
two lines reflecting the $1\sigma$ statistical uncertainties on
pressure at each radius. Uncertainties on temperature are not
included.}
\label{pressurepr}
\end{center}
\end{figure*}

If we assume that the source is in the plane of the sky (measured
lengths are true lengths) and that there is no beaming, the results
are shown in Table \ref{pressures} and plotted in Fig.\
\ref{pressurepr}. We see that the minimum pressures are close to the
external thermal pressures for all these regions, except for the lobe,
where the internal minimum pressure is much lower than the external
pressure, as seen in other sources. (Note however that the outer
pressure for the lobe is an extrapolation of the fitted $\beta$ model
well beyond the region where emission is seen, so that it should be
treated with caution.) As the effects of beaming will tend to reduce
the pressures in the jet components, these results suggest (contrary
to what was found in earlier work) that the jet can be close to
pressure balance with the external medium over its entire length. We
have repeated the calculations using two beaming models: firstly, one
that can reproduce the sidedness in the inner kiloparsec-scale jet
($\theta = 40\degr$ and $\beta = 0.8$), and secondly, one that could
produce the X-ray emission from region 1 on an inverse-Compton model
($\theta = 12\degr$, $\beta = 0.98$). For both these models, the
change in the ratio of radio to thermal pressure is uniformly a
decrease (though in the second case the thermal pressures at large
distances are uncertain). Including the possible effects of beaming
and projection thus only strengthens the conclusion that the jet is
underpressured, or at best in pressure balance with the thermal
emission at minimum pressure. The main differences between our
analysis and that of \cite{bw93} are (1) the additional small-scale
thermal component, (2) slightly lower minimum pressures in this work
[as a result of slightly more conservative assumptions than those used
by \cite{per84}] and (3) slightly higher estimated thermal pressures
in the large-scale thermal emission. We stress that our calculation
assumes that the jet is electron-positron (with no heavy content) and
is radiating at minimum energy.

\begin{figure*}
\begin{center}
\includegraphics[width=10cm]{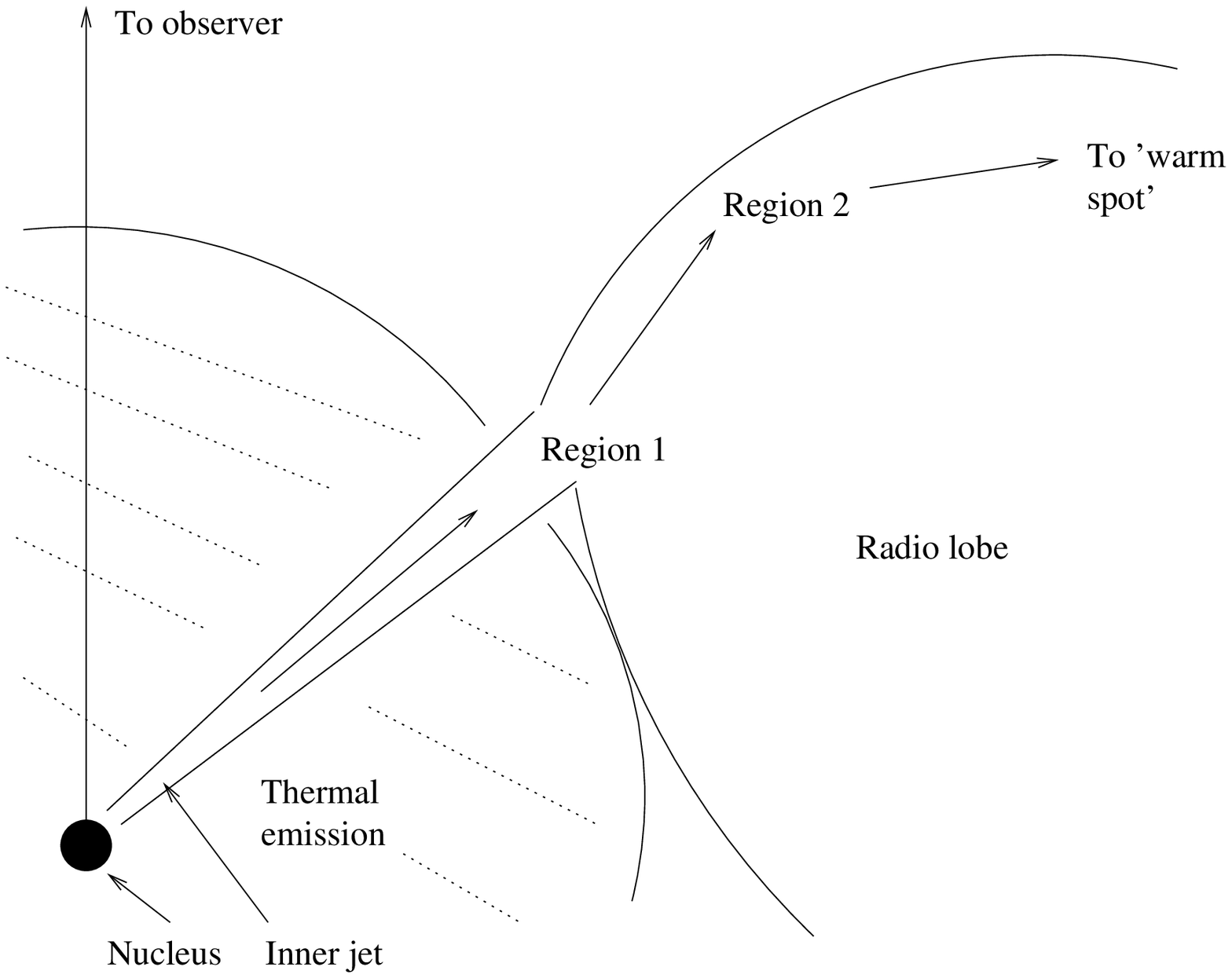}
\caption{A sketch of a possible geometry for the N jet and lobe of NGC
  6251 (not to scale). The observer is on the vertical axis, and the
  plane of the sky is perpendicular to the page. Thermal emission
  (sketched as a single contour of constant pressure; it should be
  assumed to fill the entire region) provides the pressure to confine
  the jet at small distances from the nucleus. The jet enters the lobe
  at region 1 and the change in conditions triggers particle
  acceleration, perhaps at a shock. The jet then bends at region 1 and
  interacts with the edge of the lobe at region 2 before continuing to
  feed the distant `warm spot' at the end of the lobe. The
  discontinuity between the lobe and the thermal X-ray emission
  appears relatively sharp to a distant observer in this source
  geometry.}
\label{sketch}
\end{center}
\end{figure*}

Does the extended thermal emission tell us anything about the
structure and dynamics of the radio source? We begin by noting that
Fig.~\ref{ext_overlays} appears to show an arm of X-ray-emitting gas
at the northern edge of the radio lobe, similar to the rim of material
seen around the southern lobe of 3C\,449 (\citealt{har98,cro03}). If
real, this feature would indicate that the lobe is behaving similarly
to other FRI radio galaxies and inflating a large-scale cavity in the
X-ray gas. The FOV of \XMM prevents us from seeing whether this
feature extends to the far end of the N lobe. More interestingly, from
the point of view of the overall interpretation, there is a
significant drop in surface brightness, first pointed out by
\cite{ker03}, evident in Fig.~\ref{ext_overlays} at roughly the
distance of region 1 of the X-ray jet, where the radio and X-ray jets
brighten and the extended lobe radio emission begins. (This drop in
surface brightness is not obvious in the radial profiles of Fig.\
\ref{prof}, but these specifically exclude emission in the direction
of the jet.) If the positional agreement between the edge of the radio
lobe and the drop in X-ray surface brightness is not coincidental, it
implies that the radio {\it lobe} axis is reasonably close to the
plane of the sky -- if the lobe were strongly projected, it would be
hard to see a discontinuity in the X-ray. This is not impossible to
reconcile with the constraints on jet angle to the line of sight
$\theta \la 40\degr$, since the lobe subtends a large angle
($70\degr$) as seen from the nucleus, provided that the jet enters the
lobe towards the surface nearer to the observer, though it is hard to
reconcile with the very small angles to the line of sight required in
an inverse-Compton model. If the jet X-ray emission is synchrotron,
then the fact that there is relatively strong X-ray and radio emission
at region 1 may be a reflection of particle acceleration at a pressure
discontinuity, or at some other interaction as the jet enters the
lobe. The weaker X-ray emission at region 2 could then be an
interaction between the jet and the edge of the lobe, since the jet
bends between region 1 and region 2, and must bend again if it is to
connect with the `warm spot' at the end of the lobe. The geometry of
this model of the jet and lobe is sketched in Figure~\ref{sketch}.

Finally, we note that, by integrating the best-fitting large-scale
$\beta$-model to derive the group-scale luminosity, and using the
radio-quiet $L_{\rm X}/T_{\rm X}$ relation (\citealt{cro03}), we would
predict a temperature of $\sim 0.65$ keV, which is significantly less
than the measured value. This may suggest some heating of the
thermally emitting gas by the radio source. As the radio lobes of this
source are large, the $P$d$V$ work done in expansion is expected to be
considerable.  We calculate, based on the external pressure at the
radius of the lobes, and an assumed volume of $1.5 \times 10^{66}$
m$^{3}$ (modelling the lobe as a sphere of radius 230 kpc, based on
the extent of the low-frequency radio emission), that the work done by
the western lobe is $\sim 3 \times 10^{52}$ J. A simple calculation of
the heat capacity of the entire group, $C = (3/2) Nk$, where $N$ is
the total number of particles, calculated using the spectral and
spatial parameters for the group atmosphere, shows that if the energy
transferred from the radio lobes is distributed throughout the group
atmosphere it would raise the temperature of the gas by $\sim 0.6$
keV. Therefore, if the eastern lobe also contributes a similar amount
of energy (it is of similar size), then the temperature of the
NGC~6251 group gas may be explained as a result of radio-source
heating due to expansion of the lobes. This is consistent with our
observations of other sources (\citealt{cro03,cro05}), although the
uncertainties are large.

\section{Conclusions}

We have presented results from {\it Chandra}/ACIS-S and ACIS-I and
{\it XMM-Newton}/EPIC observations of the nucleus, jet, and extended
emission of the radio galaxy NGC~6251. For the nucleus, we find the
following.

\begin{enumerate}
\item The X-ray spectrum is well fitted by an absorbed
power-law with thermal emission. There is tentative, but not highly
significant, evidence for the detection of an Fe K$\alpha$ line with
{\it{XMM-Newton}}.

\item The nuclear spectral energy distribution is double-peaked, similar to that
found by \cite{chi03}, and is well-fitted with a synchrotron
self-Compton model. This lends support to a model in which the nuclear
X-ray emission is due to inverse-Compton upscattering of low energy
photons in a relativistic jet to higher energies.

\item It is plausible that some of the X-ray emission originates 
in an accretion flow (rather than a jet), and that the relative
dominance of accretion- and jet-related components varies with time.

\end{enumerate}

For the kiloparsec-scale jet and extended emission, our conclusions
are as follows.

\begin{enumerate}

\item There is a well-detected inner X-ray jet with a broad-band
  spectrum and X-ray properties similar to those of other X-ray
  jet sources, implying local particle acceleration in the inner
  10--20 kpc.

\item Inverse-Compton models for the previously detected large-scale
  jet emission (`regions 1 and 2' in our notation) require extreme
  parameters; coincidences in geometry are required to explain the non-detection of
  the inner part of the jet. Although a one-zone synchrotron model is
  only marginally viable for region 1, we prefer a synchrotron over an
  inverse-Compton explanation for both this region and region 2. In
  the synchrotron picture, regions 1 and 2 may be privileged sites for
  particle acceleration as a result of interactions between the jet
  and the lobe.

\item Kiloparsec-scale thermal emission is present in NGC~6251, with
  properties similar to that found in other radio-loud ellipticals, as
  well as the previously known group-scale thermal emission. We are
  able to characterize the properties of both the galaxy-scale and
  group-scale components and show that the pressure of the thermal gas
  can confine the jet if it is at minimum energy and contains no
  significant contribution from heavy particles, particularly if (as
  seems likely) relativistic beaming is important.

\item The detailed relationship between the northern radio lobe and
  the extended X-ray emission suggests that the radio lobe has
  evacuated a cavity in the X-ray-emitting gas, as seen in other FRI
  sources. If so, the sharpness of the boundary between gas and lobe
  suggests that the lobe is close to the plane of the sky, and that
  the jet enters the lobe close to the surface nearer the observer.

\item The group gas is somewhat hotter than expected for its
  luminosity, and the work done in expanding the radio lobes provides
  a plausible source for the additional heat.

\end{enumerate}

\section{Acknowledgements}

We are grateful for support for this work from PPARC (Studentships for
DAE and JHC, and a research grant for DMW) and the Royal Society
(Research Fellowship for MJH). We thank Dayton Jones for providing the
VLBI data and Pierre Werner for allowing us to use results from his
Ph.D. thesis prior to publication. DAE thanks the Harvard-Smithsonian
Center for Astrophysics for its support, Ralph Kraft, Jane Turner and
Sandor Molnar for useful discussions, Masahiro Tsujimoto for the
ARF-correction software, and Alexey Vikhlinin for the {\sc zhtools}
software. We thank the anonymous referee for useful comments.


\clearpage

\clearpage
\begin{table*}
\caption{\XMM observed 0.15--10 keV count rates in a source-centred
  circle of radius 35 arcsec, together with nominal count rates above
  which pileup becomes important (as defined in the \XMM Users' Handbook)}
\label{xmm_pileup_tab}
\begin{tabular}{lccc}
\hline
Camera & Total exposure time (ks) & Count rate (cts s$^{-1}$)  & Pile up threshold (cts s$^{-1}$)\\
\hline
MOS1   & 23.7                     & 0.556 $\pm$ 0.005          & 0.7 \\
MOS2   & 24.7                     & 0.574 $\pm$ 0.005          & 0.7 \\
pn     & 8.1                      & 1.973 $\pm$ 0.016          & 8   \\
\hline
\end{tabular}
\end{table*}

\begin{table*}
\caption{VLA observations of NGC~6251 used in the analysis}
\label{vla}
\begin{tabular}{llrrrllr}
\hline
Proposal&Date&Freq.&Bandwidth&Duration&Config.&Map resn&Ref.\\
ID&&(GHz)&(MHz)&(h)&&(arcsec)\\
\hline
VJ38&1985 Apr 08&1.665&$2 \times 50$&5.0&B&$3.0\times 2.8$&1\\
AK153&1986 Aug 20&1.452&25&2.8&B&$4.9 \times 3.7$\\
AZ032&1987 Aug 28&333&$64 \times 0.05$&7.0&A&$4.0 \times 4.0$&4\\
AZ35&1987 Dec 27,&333&$64 \times 0.05$&5.0&B&$19 \times 12$\\
&1988 Jan 09\\
VJ49&1988 Nov 20&1.665&$2 \times 50$&11.6&A&$1.3 \times 1.1$&2\\
AP0001&1990 Jun 27&8.427&$7 \times 3.125$&3.3&BnA&$0.77 \times 0.58$\\
AM322&1991 May 09&8.415&$2 \times 50$&1.2&D&$8.7 \times 7.7$\\
AB746&1995 Aug 15&1.387&$32 \times 0.4$&3.0&A&$1.7 \times 1.2$&3\\
\hline
\end{tabular}
\vskip 10pt
\begin{minipage}{15.5 cm}
References are (1) \cite{jonetal86} (2) \cite{jon94} (3) \cite{wer02} and Werner \etal\ (in prep.) (4) Birkinshaw \etal, in
prep. The resolutions quoted are major axis $\times$ minor axis (FWHM)
of the elliptical Gaussian restoring beam. The duration quoted is the
time spent on source.
\end{minipage}
\end{table*}

\begin{table*}
\caption{\Ch and \XMM best-fitting spectral parameters}
\label{chandra_xmm_spectral_tab}
\begin{tabular}{lccccc}
\hline
Observation & Inner Extraction Radius & Outer Extraction Radius & Component & Parameters & Normalization\\
\hline
{\it Chandra}/ACIS-S & 0.492 arcsec & 1.23 arcsec & Power Law & $\Gamma$ = 1.67 $\pm$ 0.06                                 & ($9.28^{+0.36}_{-0.33}$)$\times$10\(^{-4}\)\\
           &              &             & Thermal   & kT = 0.20 $\pm$ 0.08 keV                                           & (1.04 $\pm$ 1.22)$\times$10\(^{-4}\) \\
           &              &             &           & Abundance = 0.35 (frozen)                                          & \\
           & 0.492 arcsec & 35 arcsec   & Power Law & $\Gamma$ = 1.69 $\pm$ 0.09                                         & (9.71 $\pm$ 0.37)$\times$10\(^{-4}\) \\
           &              &             &           &                                                                    & \\
           &              &             & Thermal   & kT = 0.59 $\pm$ 0.05 keV                                           & (1.81 $\pm$ 0.26)$\times$10\(^{-4}\) \\
           &              &             &           & Abundance = 0.35 (frozen)                                          & \\
      \XMM & 0 arcsec     & 35 arcsec   & Power Law & $\Gamma$ = 1.88 $\pm$ 0.01                                         & MOS1 = (1.20 $\pm$ 0.02)$\times$10\(^{-3}\) \\ 
           &              &             &           &                                                                    & MOS2 = (1.22 $\pm$ 0.02)$\times$10\(^{-3}\) \\
           &              &             &           &                                                                    & pn = (1.27 $\pm$ 0.02)$\times$10\(^{-3}\) \\
           &              &             & Thermal   & kT = 0.58 $\pm$ 0.03 keV                                           & MOS1 = (4.87 $\pm$ 3.93)$\times$10\(^{-5}\) \\
           &              &             &           & Abundance = 0.35 (frozen)                                          & MOS2 = (6.41 $\pm$ 4.06)$\times$10\(^{-5}\) \\
           &              &             &           &                                                                    & pn = (2.07 $\pm$ 0.28)$\times$10\(^{-4}\) \\
\hline
\end{tabular}
\vskip 10pt
\begin{minipage}{15.5 cm}
The thermal abundance was fixed at 0.35 solar, and the intrinsic
absorption (in excess of the Galactic value) fixed at $4.50 \times
10^{20}$ atoms cm$^{-2}$. Uncertainties are 90 per cent for one
parameter  of interest (i.e., $\chi^2_{\rm min}$ + 2.7). The \XMM
parameters quoted are for the 3$\sigma$ GTI-filtering method defined
in Section 3.2. The parameters derived with the other \XMM GTI-filtering
methods are consistent with those quoted here.
\end{minipage}
\end{table*}

\begin{table*}
\caption{Summary of \XMM Exposures and Counts After GTI-Filtering}
\label{exposures_tab}
\begin{scriptsize}
\begin{tabular}{lccccccccc}
\hline
                           & \multicolumn{3}{c}{MOS1} & \multicolumn{3}{c}{MOS2} & \multicolumn{3}{c}{pn}   \\
GTI filtering applied      &          & 2--10 keV     & 5.5--7.5 keV  &          & 2--10 keV     & 5.5--7.5 keV  &          & 2--10 keV     & 5.5--7.5 keV \\
                           & Exposure & Source Counts & Source Counts & Exposure & Source Counts & Source Counts & Exposure & Source Counts & Source Counts \\
\hline
Conservative                              & 26.4 ks  &  4,041 & 325        & 27.5 ks  &  4,041 & 330        & 15.8 ks  &  6,861 & 868        \\
3$\sigma$ (from high-energy light curves) & 43.2 ks  &  6,670 & 553        & 43.0 ks  &  6,472 & 601        & 38.1 ks  & 18,216 & 2503       \\
3$\sigma$ (from low-energy light curves)  & 43.9 ks  &  6,789 & 544        & 43.7 ks  &  6,599 & 627        & 39.5 ks  & 19,179 & 2701       \\
\hline
\end{tabular}
\end{scriptsize}
\end{table*}

\begin{table*}
\caption{Significance of \XMM Fe-line addition}
\label{fe_sig_tab}
\begin{tabular}{lccc}
\hline
GTI filtering applied & MOS1+MOS2 & pn & MOS1+MOS2+pn \\
\hline
Conservative & 100\% & 16.9\% & 87.2\% \\
3$\sigma$ (from high-energy light curves) & 100\% & 17.0\% & 36.0\% \\
3$\sigma$ (from low-energy light curves) & 100\% & 14.8\% & 9.0\% \\
\hline
\end{tabular}
\vskip 10pt
\begin{minipage}{15.5 cm}
Percentages quoted are the probability of achieving a greater $F$ by chance.
\end{minipage}
\end{table*}

\begin{table*}
\caption{Key spectral parameters after fitting Fe K$\alpha$ line to the \XMM data}
\label{fe_results_tab}
\begin{scriptsize}
\begin{tabular}{llll}
\hline
GTI filtering applied & MOS1+MOS2 & pn & MOS1+MOS2+pn \\
\hline
Conservative                 & $\Gamma$ = $1.77^{+0.15}_{-0.07}$ & $\Gamma$ = $1.95^{+0.17}_{-0.12}$                & $\Gamma$ = $1.87^{+0.13}_{-0.08}$           \\
                             & E = N/A                           & E = 6.4 keV (pegged)                             & E = 6.51 keV (pegged) \\
                             & $\sigma$ = N/A                    & $\sigma = 0.76 \pm 0.59 $ keV                    & $\sigma = 4.02 \times 10^{-8}$ keV ($\pm$ unconstrained) \\
                             & norm = 0                          & norm = ($7.61^{+9.42}_{-7.61}) \times 10^{-6}$   & norm = 6.52 $\times 10^{-7}$ ($\pm$ unconstrained) \\
3$\sigma$ (from high-energy  & $\Gamma$ = $1.84^{+0.16}_{-0.09}$ & $\Gamma$ = $1.86^{+0.11}_{-0.04}$                & $\Gamma$ = $1.85^{+0.09}_{-0.04}$           \\
light curves)                & E = N/A                           & E = $6.83^{+0.95}_{-0.62}$ keV                   & E = $6.90 \pm 3.29$ keV \\
                             & $\sigma$ = N/A                    & $\sigma$ = $0.56^{+0.94}_{-0.54}$ keV            & $\sigma$ = $0.51^{+1.93}_{-0.51}$ keV\\
                             & norm = 0                          & norm = ($4.46^{+5.02}_{-4.46} \times 10^{-6}$)   & norm = ($2.45^{+2.69}_{-2.45} \times 10^{-6}$)\\
3$\sigma$ (from low-energy   & $\Gamma$ = $1.84^{+0.16}_{-0.09}$ & $\Gamma$ = $1.87^{+0.11}_{-0.05}$                & $\Gamma$ = $1.85^{+0.09}_{-0.04}$           \\
light curves)                & E = N/A                           & E = $6.49 \pm 0.64)$                             & E = $6.46^{0.13}_{-0.45}$ \\
                             & $\sigma$ = N/A                    & $\sigma$ = $0.38 \pm 0.26$                       & $\sigma$ = $1.77^{+9.71}_{-1.77} \times 10^{-2}$\\
                             & norm = 0                          & norm = ($3.98^{+6.02}_{-3.98} \times 10^{-6}$)   & norm = ($1.60^{+1.04}_{-1.60} \times 10^{-6}$)\\ \hline
\end{tabular}
\end{scriptsize}
\vskip 10pt
\begin{minipage}{15.5 cm}
Uncertainties are 90 per cent for one interesting parameter (i.e.,
$\chi^2_{\rm min}$ + 2.7).
\end{minipage}
\end{table*}

\begin{table*}
\caption{$\chi^2$ results for fitting a constant count rate to the
  \XMM and \Ch data sets with 3,000 second time bins. The \XMM GTI
  filtering was identical to that of \cite{gli04}.}
\label{chandra_xmm_vartab}
\begin{tabular}{llccccc}
\hline
Observatory & Camera       & Energy range (keV) & Average Count Rate (cts s$^{-1}$) & $\chi^2$ / dof & Null hypothesis probability\\
\hline
\XMM        & MOS1         & 0.8--10            & 0.42                              & 8.45/13                 & 0.82                        \\
            & MOS2         &                    & 0.43                              & 15.29/13                & 0.29                        \\
            & pn           &                    & 1.17                              & 22.35/13                & 0.05                        \\
            & MOS1+MOS2+pn &                    & 2.03                              & 26.07/12                & 0.01                        \\
\Ch         & ACIS-S       & 0.3--10            & 0.68                              & 13.37/16                & 0.65                        \\
\hline
\end{tabular}
\end{table*}

\begin{table*}
\caption{Long-term X-ray variability history of NGC~6251}
\label{longterm_vartab}
\begin{tabular}{llcccl}
\hline
Date            & Observatory       & 2--10 keV PL-only luminosity        & Photon Index           & Reference     \\
\hline
1991 March      & {\it ROSAT}       & (1.8 $\pm$ 0.5) $\times$10\(^{42}\) & 2.0$^{+0.6}_{-0.5}$    & \cite{bw93}   \\
1994 October    & {\it ASCA}        & (1.7 $\pm$ 0.1) $\times$10\(^{42}\) & 2.11$^{+0.16}_{-0.19}$ & \cite{tur97}  \\
2000 September  & \Ch               & (4.3 $\pm$ 0.4) $\times$10\(^{42}\) & 1.76 $\pm$ 0.16        & \cite{gua03}  \\
2001 July       & {\it BeppoSAX}    & (7.1 $\pm$ 0.6) $\times$10\(^{42}\) & 1.70$^{+0.12}_{-0.16}$ & \cite{gua03}  \\
2002 March      & \XMM              & (5.1 $\pm$ 0.1) $\times$10\(^{42}\) & 1.88 $\pm$ 0.01        & Present work  \\
2003 November   & \Ch               & (6.1 $\pm$ 0.2) $\times$10\(^{42}\) & 1.67 $\pm$ 0.08        & Present work  \\
\hline
\end{tabular}
\vskip 10pt
\begin{minipage}{15.5 cm}
All luminosities quoted are unabsorbed. For {\it ROSAT}, the observed spectrum has been extrapolated to the 2--10 keV energy range. The errors quoted are statistical, i.e., 1$\sigma$.
\end{minipage}
\end{table*}

\begin{table*}
\caption{VLBI flux density estimates for the radio core and parsec-scale jet
  at several epochs and frequencies}
\label{vlbi_vartab}
\begin{tabular}{cccccc}
\hline
Frequency & Date & Core flux density (Jy) & Jet flux density (Jy) & Total flux density (Jy) & Reference\\
\hline
1.6 GHz & 1983 March    & 0.29            & 0.24            & 0.53            & \cite{jonetal86} \\
        & 1985 April    & 0.21 $\pm$ 0.03 & 0.12 $\pm$ 0.02 & 0.33 $\pm$ 0.04 & Present work \\
        & 1988 November & 0.42 $\pm$ 0.03 & 0.09 $\pm$ 0.01 & 0.51 $\pm$ 0.04 & Present work \\
        & 2000 May      & 0.31 $\pm$ 0.02 & 0.10 $\pm$ 0.01 & 0.41 $\pm$ 0.02 & Present work \\
2.3 GHz & 1980 February & 0.25            & 0.40            & 0.65            & \cite{jonetal86} \\
5.0 GHz & 1981 December & 0.35            & 0.54            & 0.89            & \cite{jonetal86} \\
        & 2000 May      & 0.38 $\pm$ 0.04 & 0.13 $\pm$ 0.01 & 0.51 $\pm$ 0.05 & Present work \\
8.4 GHz & 2000 May      & 0.38 $\pm$ 0.04 & 0.13 $\pm$ 0.01 & 0.51 $\pm$ 0.05 & Present work \\
10.7 GHz& 1978 May      & 0.59            & 0.30            & 0.89            & \cite{coh79}, \cite{jonetal86} \\
15 GHz  & 2000 May      & 0.39 $\pm$ 0.04 & 0.11 $\pm$ 0.01 & 0.50 $\pm$ 0.05 & Present work \\
\hline
\end{tabular}
\end{table*}

\begin{table*}
\caption{Multi-epoch archival {\it HST} nuclear flux densities}
\label{hst_vartab}
\begin{tabular}{lllccc}
\hline
Instrument & Filter & Date & Exposure time (seconds) & Pivot wavelength (\AA) & Flux density ($\mu$Jy)\\
\hline
WFPC2 & F555W & 1994 February 21  & 70      & 5442 & 171.56 $\pm$ 5.82\\
WFPC2 & F555W & 1995 June 28      & 800     & 5442 & 182.18 $\pm$ 1.77\\
WFPC2 & F814W & 1995 June 28      & 720     & 8002 & 244.95 $\pm$ 2.66\\
WFPC2 & F814W & 1996 September 13 & 1000    & 8002 & 248.38 $\pm$ 2.27\\
FOC   & F342W & 1991 May 12       & 1196    & 3403 & 34.56  $\pm$ 0.41\\
FOC   & F342W & 1993 September 30 & 2246    & 3403 & 19.36  $\pm$ 0.22\\
FOC   & F342W & 1996 February 19  & 1196    & 3403 & 54.64  $\pm$ 0.45\\
FOC   & F410M & 1993 September 30 & 1676    & 4088 & 26.70  $\pm$ 0.80\\
FOC   & F410M & 1996 February 19  & 1315    & 4088 & 91.22  $\pm$ 1.34\\
\hline
\end{tabular}
\vskip 10pt
\begin{minipage}{15.5 cm}
The quoted values have been dereddened assuming that the optical extinction to NGC~6251 is the sum of both the measured Galactic and intrinsic absorptions
\end{minipage}
\end{table*}

\begin{table*}
\caption{X-ray and radio flux densities and ratios for the kiloparsec-scale and
parsec-scale jets}
\label{fluxratios_tab}
\begin{tabular}{lccc}
\hline
Component                     & 1 keV X-ray flux density (Jy) & Radio flux density (Jy) (frequency) & Ratio ($\times$10\(^{-6}\))   \\
\hline
Inner (kiloparsec-scale) jet  & 1$\times$10\(^{-8}\)          & 3.1$\times$10\(^{-2}\) (8 GHz)      & 0.32                          \\
Core power law / pc-scale jet & 6.2$\times$10\(^{-7}\)        & $\sim$ 1 (5 GHz)                    & 0.62                          \\
\hline
\end{tabular}
\end{table*}

\begin{table*}
\caption{X-ray-to-radio flux density ratios for a selection of FRI type radio galaxies}
\label{jetxrflux}
\begin{tabular}{lcl}
\hline
Source  & 1-keV X-ray to 8-GHz radio                    & Reference    \\
        & flux density ratio ($\times$10\(^{-6}\))      &              \\
\hline
NGC~6251& 0.32                                          & Present Work \\
3C 66B  & 0.13                                          & \cite{har01} \\
3C 31   & 0.24                                          & \cite{har02b}\\
Cen A   & 0.07--9.7 (various knots)                     & \cite{har03} \\
NGC 315 & 0.083                                         & \cite{wor03} \\
3C 296  & 0.069                                         & \cite{har05} \\
\hline
\end{tabular}
\begin{minipage}{15cm}
The X-ray and radio measurements are over the X-ray-detected regions of the jet.
\end{minipage}
\end{table*}

\begin{table*}
\caption{Minimum and thermal pressures in radio components of NGC
  6251}
\label{pressures}
\begin{tabular}{lrlrr}
\hline
Region&Minimum pressure&Range&Inner $p$&Outer $p$\\
&(Pa)&(arcsec)&(Pa)&(Pa)\\
\hline
Inner jet (optical)&$9.7 \times 10^{-13}$&8--20&$4.1^{+0.3}_{-0.2}
\times 10^{-12}$&$11.0^{+0.9}_{-0.6} \times 10^{-13}$\\
Middle jet&$1.5 \times 10^{-13}$&30--200&$7.8^{+0.9}_{-0.8} \times 10^{-13}$&$1.3^{+0.1}_{-0.1} \times 10^{-13}$\\
Outer jet region 1&$1.4 \times 10^{-13}$&200--265&$1.3^{+0.1}_{-0.1} \times 10^{-13}$&$7.2^{+1.4}_{-1.5} \times 10^{-14}$\\
Outer jet region 2&$3.2 \times 10^{-14}$&330--410&$4.3^{+1.4}_{-1.6} \times 10^{-14}$&$2.5^{+1.2}_{-1.3} \times 10^{-14}$\\
N lobe&$3.2 \times 10^{-16}$&235--1000&$9.4^{+1.3}_{-1.4} \times 10^{-14}$&$2.5^{+3.9}_{-2.2} \times 10^{-15}$\\
\hline
\end{tabular}
\vskip 10pt
\begin{minipage}{15cm}
Pressure for inner jet is calculated from the count density from the
combination of the small- and large-scale $\beta$-model fits, assuming
$kT = 0.59$ keV. Other pressures are calculated using the large-scale
component only, assuming $kT = 1.7$ keV.
\end{minipage}
\end{table*}


\begin{thebibliography}{99}

  \bibitem[\protect\citeauthoryear{Arnaud et al.}{2002}]{arn02} Arnaud M., et al., 2002, A\&A, 390, 27

  \bibitem[\protect\citeauthoryear{Birkinshaw \& Worrall}{1993}]{bw93} Birkinshaw, M., Worrall, D. M., 1993, ApJ, 412, 568

  \bibitem[\protect\citeauthoryear{Brunetti}{2000}]{bru00} Brunetti G., 2000, Astroparticle Physics, 13, 107

  \bibitem[\protect\citeauthoryear{Burstein \& Heiles}{1978}]{bur78} Burstein D., Heiles C., 1978, ApJ, 225, 40

  \bibitem[\protect\citeauthoryear{Canosa et al.}{1999}]{can99} Canosa C.~M., Worrall D.~M., Hardcastle M.~J., Birkinshaw M., 1999, MNRAS, 310, 30 

  \bibitem[\protect\citeauthoryear{Chiaberge et al.}{2000}]{chi00} Chiaberge M., Celotti A., Capetti A., Ghisellini G. 2000, A\&A, 358, 104

  \bibitem[\protect\citeauthoryear{Chiaberge et al.}{2003}]{chi03} Chiaberge, M., Gilli, R., Capetti, A., Macchetto, F. D. 2003, ApJ, 597, 166 

  \bibitem[\protect\citeauthoryear{Cohen \& Readhead}{1979}]{coh79} Cohen, M. H., Readhead, A. C. S. 1979, ApJ, 233, L101

  \bibitem[\protect\citeauthoryear{Croston et al.}{2003}]{cro03} Croston J. H., Hardcastle M. J., Birkinshaw M., Worrall D. M., 2003, MNRAS, 346, 1041

  \bibitem[\protect\citeauthoryear{Croston et al.}{2004}]{cro04a} Croston J.~H., Birkinshaw M., Hardcastle M.~J., Worrall D.~M., 2004, MNRAS, 353, 879
 
  \bibitem[\protect\citeauthoryear{Croston, Hardcastle, \& Birkinshaw}{2005}]{cro05} Croston J.~H., Hardcastle M.~J., Birkinshaw M., 2005, MNRAS, 357, 279

  \bibitem[\protect\citeauthoryear{Davis}{2001}]{dav01} Davis, J. E. 2001, ApJ, 562, 575

  \bibitem[\protect\citeauthoryear{Evans et al.}{2004}]{evans04} Evans D. A., Kraft R. P., Worrall D. M., Hardcastle M. J., Jones C., Forman W. R., Murray S. S., 2004, ApJ, 612, 786 

  \bibitem[\protect\citeauthoryear{Fabbiano et al.}{1984}]{fab84} Fabbiano, G., Trinchieri, G., Elvis, M., Miller, L., Longair, M. 1984, ApJ, 277, 115

  \bibitem[\protect\citeauthoryear{Ferrarese \& Ford}{1999}]{fer99} Ferrarese, L., Ford, H. C. 1999, ApJ, 515, 583 

  \bibitem[\protect\citeauthoryear{Foschini et al.}{2004}]{fos04} Foschini L., et al., 2004, A\&A accepted, preprint (astro-ph/0412285)

  \bibitem[\protect\citeauthoryear{Gambill et al.}{2003}]{gam03} Gambill, J. K., Sambruna, R. M., Chartas, G., Cheung, C. C., Maraschi, L., Tavecchio, F., Urry, C. M., Pesce, J. E. 2003, A\&A, 401, 505 

  \bibitem[\protect\citeauthoryear{Ghisellini, Tavecchio, \& Chiaberge}{2004}]{ghi04} Ghisellini, G., Tavecchio, F., Chiaberge, M. 2004, A\&A accepted, preprint (astro-ph/0406093)

  \bibitem[\protect\citeauthoryear{Ghizzardi et al.}{2001b}]{ghi01b} Ghizzardi, S. 2001, ``In-flight calibration of the PSF for the PN camera'', XMM-SOC-CAL-TN-0023, available from http://www.xmm.vilspa.esa.es
  
  \bibitem[\protect\citeauthoryear{Gliozzi et al.}{2004}]{gli04} Gliozzi, M., Sambruna, R. M., Brant, W. N., Mushotzky, R., Eracleous, M. 2004, A\&A, 413, 139

  \bibitem[\protect\citeauthoryear{Guainazzi et al.}{2003}]{gua03} Guainazzi, M., Grandi, P., Comastri, A., Matt, G. 2003, A\&A, 410, 131 

  \bibitem[\protect\citeauthoryear{Hardcastle, Worrall, \& Birkinshaw}{1998}]{har98} Hardcastle M.~J., Worrall D.~M., Birkinshaw M., 1998, MNRAS, 296, 1098

  \bibitem[\protect\citeauthoryear{Hardcastle \& Worrall}{1999}]{har99} Hardcastle, M. J., Worrall, D. M. 1999, MNRAS, 309, 969

  \bibitem[\protect\citeauthoryear{Hardcastle \& Worrall}{2000}]{har00} Hardcastle M.~J., Worrall D.~M., 2000, MNRAS, 314, 359

  \bibitem[\protect\citeauthoryear{Hardcastle, Birkinshaw, \& Worrall}{2001}]{har01} Hardcastle M. J., Birkinshaw M., Worrall D. M., 2001, MNRAS, 326, 1499

  \bibitem[\protect\citeauthoryear{Hardcastle et al.}{2002a}]{har02a} Hardcastle M. J., Birkinshaw M., Cameron R. A., Harris D. E., Looney L. W., Worrall D. M., 2002, ApJ, 581, 948

  \bibitem[\protect\citeauthoryear{Hardcastle et al.}{2002b}]{har02b} Hardcastle, M. J., Worrall, D. M., Birkinshaw, M., Laing, R. A., Bridle, A. H. 2002, MNRAS, 334, 182 

  \bibitem[\protect\citeauthoryear{Hardcastle et al.}{2003}]{har03} Hardcastle M.~J., Worrall D.~M., Kraft R.~P., Forman W.~R., Jones C., Murray S.~S., 2003, ApJ, 593, 169

  \bibitem[\protect\citeauthoryear{Hardcastle et al.}{2005}]{har05} Hardcastle, M. J., Worrall, D. M., Birkinshaw, M., Laing, R. A., Bridle, A. H. 2005, MNRAS, accepted, preprint (astro-ph/0412599)

  \bibitem[\protect\citeauthoryear{Ho}{1999}]{ho99} Ho L.~C., 1999, ApJ, 516, 672
  
  \bibitem[\protect\citeauthoryear{Jones et al.}{1986}]{jonetal86} Jones, D. L., et al. 1986, ApJ, 305, 684 

  \bibitem[\protect\citeauthoryear{Jones \& Wehrle}{1994}]{jon94} Jones D. L., Wehrle A. E., 1994, ApJ, 427, 221

  \bibitem[\protect\citeauthoryear{Jones \& Wehrle}{2002}]{jon02} Jones, D. L., Wehrle, A. E. 2002, ApJ, 580, 114 

  \bibitem[\protect\citeauthoryear{Kerp \& Mack}{2003}]{ker03} Kerp, J., Mack, K.-H. 2003, New Astronomy Review, 47, 447 

  \bibitem[\protect\citeauthoryear{Laing et al.}{1999}]{lai99} Laing R. A., Parma P., de Ruiter H. R., Fanti R., 1999, MNRAS, 306, 513

  \bibitem[\protect\citeauthoryear{Laing \& Bridle}{2004}]{lai04} Laing R. A., Bridle A. H., 2004, MNRAS, 348, 1459

  \bibitem[\protect\citeauthoryear{Mack, Kerp, \& Klein}{1997a}]{mac97a} Mack K.-H., Kerp J., Klein U., 1997, A\&A, 324, 870

  \bibitem[\protect\citeauthoryear{Mack et al.}{1997b}]{mac97b} Mack K.-H., Klein U., O'Dea C. P., Willis A. G., 1997, A\&AS, 123, 423

  \bibitem[\protect\citeauthoryear{Marshall et al.}{2004}]{mar04} Marshall, H. L., Tennant, A., Grant, C. E., Hitchcock, A. P., O'Dell, S. L., Plucinsky, P. P. 2004, SPIE, 5165, 497

  \bibitem[\protect\citeauthoryear{Marshall et al.}{2005}]{mar05} Marshall, H. L., {\emph{et al.}} 2005, ApJS 156, 13

  \bibitem[\protect\citeauthoryear{Merloni, Heinz, \& Di Matteo}{2003}]{mer03} Merloni A., Heinz S., Di Matteo T. 2003, MNRAS, 345, 1057 

  \bibitem[\protect\citeauthoryear{Molendi \& Sembay }{2003}]{mol03} Molendi, S., Sembay, S. 2003, ``Assessing the EPIC Spectral Calibration in the Hard Band with a 3C273 Observation'', XMM-SOC-CAL-TN-0036, available from http://www.xmm.vilspa.esa.es

  \bibitem[\protect\citeauthoryear{Mukherjee et al.}{2002}]{muk02} Mukherjee, R., Halpern, J., Mirabal, N., Gotthelf, E. V. 2002, ApJ, 574, 693 

  \bibitem[\protect\citeauthoryear{Murphy et al.}{1996}]{mur96} Murphy, E. M., Lockman, F. J., Laor, A., Elvis, M. 1996, ApJS, 105, 369 

  \bibitem[\protect\citeauthoryear{Perley, Bridle, \& Willis}{1984}]{per84} Perley R. A., Bridle A. H., Willis A. G., 1984, ApJS, 54, 291

  \bibitem[\protect\citeauthoryear{Protassov et al.}{2002}]{pro02} Protassov, R., van  Dyk, D. A., Connors, A., Kashyap, V. L., Siemiginowska, A. 2002, ApJ, 571, 545 

  \bibitem[\protect\citeauthoryear{Read \& Ponman}{2003}]{rea03} Read A. M., Ponman T. J., 2003, A\&A, 409, 395

  \bibitem[\protect\citeauthoryear{Sambruna et al.}{2004}]{sam04} Sambruna, R. M., Gliozzi, M., Donato, D., Tavecchio, F., Cheung, C. C., Mushotzky, R. F. 2004, A\&A, 414, 885 

  \bibitem[\protect\citeauthoryear{Schild}{1977}]{sch77} Schild R.~E., 1977, AJ, 82, 337 

  \bibitem[\protect\citeauthoryear{Schlegel, Finkbeiner, \& Davis}{1998}]{sfd98} Schlegel D.~J., Finkbeiner D.~P., Davis M., 1998, ApJ, 500, 525

  \bibitem[\protect\citeauthoryear{Shuder \& Osterbrock}{1981}]{shu81} Shuder J.~M., Osterbrock D.~E., 1981, ApJ, 250, 55

  \bibitem[\protect\citeauthoryear{Sudou et al.}{2000}]{sud00} Sudou, H., Taniguchi, Y., Ohyama, Y., Kameno, S., Sawada-Satoh, S., Inoue, M., Kaburaki, O., Sasao, T. 2000, PASJ, 52, 989 

  \bibitem[\protect\citeauthoryear{Tavecchio et al.}{2000}]{tav00} Tavecchio F., Maraschi L., Sambruna R.~M., Urry C.~M., 2000, ApJ, 544, L23

  \bibitem[\protect\citeauthoryear{Turner et al.}{1997}]{tur97} Turner, T. J., George, I. M., Nandra, K., Mushotzky, R. F. 1997, ApJS, 113, 23

  \bibitem[\protect\citeauthoryear{Waggett, Warner, \& Baldwin}{1977}]{wag77} Waggett P. C., Warner P. J., Baldwin J. E., 1977, MNRAS, 181, 465

  \bibitem[\protect\citeauthoryear{Wilson \& Yang}{2002}]{wil02} Wilson, A. S., Yang, Y. 2002, ApJ, 568, 133

  \bibitem[\protect\citeauthoryear{Werner}{2002}]{wer02} Werner, P. N., Ph.D. thesis, Univ. Bristol

  \bibitem[\protect\citeauthoryear{Worrall \& Birkinshaw}{1994}]{wor94} Worrall, D. M., Birkinshaw, M. 1994, Ap.J., 427, 134

  \bibitem[\protect\citeauthoryear{Worrall, Birkinshaw, \& Hardcastle}{2003}]{wor03} Worrall D.~M., Birkinshaw M., Hardcastle M.~J., 2003, MNRAS, 343, L73 

\end{thebibliography}
\end{document}